\documentclass[fleqn,10pt]{wlscirep}
\usepackage[utf8]{inputenc}
\usepackage[T1]{fontenc}

\title{An Experiment with Electric Guitar Signals for Exploring the Virtuosity based on the Entropy of Music}

\author[1,*,+]{Igor Lugo}
\author[2,+]{Martha G. Alatriste-Contreras}
\affil[1]{National Autonomous University of Mexico, Centro Regional de Investigaciones Multidisciplinarias (CRIM), Cuernavaca, Mor., 62210, Mexico}
\affil[2]{National Autonomous University of Mexico, Facultad de Econom\'{i}a, Mexico City, 04510, Mexico}

\affil[*]{igorlugo@crim.unam.mx}

\affil[+]{these authors contributed equally to this work}

\begin{abstract}
We analyze the concept of virtuosity as a collective attribute in music and its relationship with the entropy based on an experiment that compares two sets of digital signals played by composer-performer electric guitarists. Based on an interdisciplinary approach related to the complex systems, we computed the spectrum of signals, identified statistical distributions that best describe them, and measured the Shannon entropy to establish their diversity. Findings suggested that virtuosity might be related to a range of entropy values that identify levels of diversity of the frequency components of audio signals. Despite the presence of different values of entropy in the two sets of signals, they are statistically similar. Therefore, entropy values can be interpreted as levels of virtuosity in music.
\end{abstract}
\begin{document}

\flushbottom
\maketitle
\section*{Introduction}

The virtuosity of playing a musical instrument is a concept most frequently related to different and subjective terms, for example ``legend,'' ``genius,'' or ``mastery.'' They describe outstanding or extraordinary abilities of musicians for composing or performing~\cite{Brossard1703,Wiener1973,Howard1993,Cooper2013}.
However, such descriptions show a lack of rationale and clarity when using them for classifying musicians. For example, through some periods in musical history, the virtuosity was defined based on the composition or performance ability. In the Renaissance period (around 1470 and 1520 CE), the term was related to the ability of creation, not only in music but also in poetry and visual arts~\cite{Beck1884}. In particular, this period was characterized by the use of a method to assist conducting accurate descriptions of different processes of the human mind. On the other hand, in the contemporary period, the virtuosity was defined as extraordinary skills and technical abilities on musical execution, for example speed of execution or precision~\cite{VanderHamm2018}. This definition identifies the musician with only one aspect of the musical practice. Then, both approaches have described the virtuosity based on the analysis of isolated individuals and on an intuitive perception of the listener.  

Nowadays, the current digital technology associated with music can explore this concept objectively and measure it based on a scientific interdisciplinary approach. In particular, the complex systems framework is one of the best approaches for analyzing collective behaviors and for connecting the musical, physical, and computational fields~\cite{Febres2017,delRioCocho2009,GonzalezEspinozaetal,GonzalezEspinozaetal2020}. Then, using digital signals, we explore the virtuosity term as the collection of waveforms that are the audio expression of the composing and performing abilities. From physical abilities to cognitive processes, the musician can generate extraordinary and memorable pieces of music. However, the identification and measurement of the patterns from which virtuosity is expressed in musical elements--for example, melody and harmony---are not trivial. Most of the time, common people misidentify the virtuoso with musicians who perform greatly due to the subjective approximation of the listener~\cite{Helmholtz1912,McLachlanetal2013,Madsenetal2019,Chanetal2019}. On the other hand, the musicians' community, who is educated and trained in the art of music, can identify more accurately physical and intellectual abilities in performing and composing music~\cite{Olsen1967,Heller2013,Grimes2014,RS2020}, but the scientific approximation to test the hypothesis of such gifted individuals is limited~\cite{BarIllan1968,Reynaud2003,Stacho2018,TG2021}. Therefore, based on the current technology related to the processing of digital signals by different devices and the complex systems approach, we study the virtuosity term by exploring a collection of data related to the fundamental property of the sound---the acoustic or mechanical wave.

At the moment, and because of the COVID-19 pandemic, musicians are using personal computers to work on different types of activities, from recording one instrument to audio mastering. The basic data collected in this type of sessions is the waveform. It describes the amplitude of the signal---higher amplitude, louder sound and lower amplitude, quieter sound~\cite{Carter2016}. To analyze this waveform, it is common to use the spectrum---the range of frequencies (cycles per second or Hertz, Hz) and their amplitude---of the audio signal
~\cite{Downey2016}. This spectrum is the result of the spectral decomposition, a procedure for simplifying data, based on the Fast Fourier Transformation (FFT) algorithm and the Discrete Fourier Transformation (DFT)~\cite{Cooleyetal1965,Press2007,SciPy2021}. Such a transformation decomposes a continuous signal into a set of periodic components showing an approximation for identifying the frequency components related to pitches and the dominant pitch and its harmonics. The type of signal we choose in our analysis is related to the electric guitar. However, it is important to mention that the electric guitar represents one instrument of a large set of musical instruments that we could select. It is beyond the scope of this study to examine all the instruments and the musical styles and players. Then, the electric guitar is particularly notable because quite a few guitarists have developed its construction and improvement, as well as its flexibility and versatility when playing it. This is evident in the case of Edward Van Halen (former musician and co-founder of the band Van Halen) because he transformed and innovated the way of composing music and playing this instrument~\cite{Waksman2001,Waksman2004, lemelson2022}.

The purpose of this study is to explore the possibility of measuring and detecting the notion of virtuosity by using a collection of digital signals related to electric guitar players. However, the selection of this data is one of the greatest challenges because the criteria for considering some levels of virtuosity are not trivial. To date there has been little agreement on which electric guitarists are considered above all others---the best of all times---due to the lack of an objective musical criterion. Therefore, for the purposes of simplicity, we consider as a virtuoso guitarists who play and compose music, from beginner to expert guitarists. We assume that the age of the guitarist is relevant for considering years of experience. Then, we generated two groups of audio signals of guitarists and compared these with each other. One set of musicians is related to those who have being considered legends by other musicians and people related to the music industry~\cite{RS2020,TG2021}; and the other set is associated with those musicians not considered as legends by the same panel of people, but they are respected guitarists in their country and in some types of social networks. We used the list of the greatest guitarists of all times provided by the Rolling Stone (RS) magazine~\cite{RS2020} as a condition for selecting a number of guitarists in the first group. In the second group, we relax this condition considering guitarists based on a wide range of levels of composer-performer guitarist (see the Materials section for the description of the experimental design). Based on this data, we applied the concept of universality and used the measure of entropy in complex systems~\cite{BarYam2004,Allenetal2017}. The universality describes a large-scale attribute of the signal even though their components differ in detail. In particular, we investigate the relationship between the variability of frequencies and their amplitudes and the statistical distribution associated with it. The measure of entropy suggested the identification of a diversity level based on the Shannon entropy~\cite{Shannon1948} as a measure of uncertainty or variability. This type of entropy measures the amount of information contained in the audio signal and determines the presence of different values, from lower to higher values---i.e., likely and unlikely frequencies. Therefore, we use an explorative data analysis searching for the possibility of identifying relationships and differences between both groups of audio signals. 

With respect to our research questions, we are interested in answering the following: 
Can the virtuosity be related to entropy in music? Does the value of entropy obviously change for different level of virtuosity? What type of statistical distributions are related to a range of entropy values?
We hypothesize that virtuosity is an emergent property of the music played with different instruments and that to be appreciated if we compare collectively similar musicians with their instruments. Compared with the virtuosity as a subjective and isolated interpretation of the composer-performer musician, we are interested in measuring and identifying the virtuosity as a range of entropy values in which at opposite ends of this range are located the noise---unstable and unclear frequencies---and the music---stable and clear frequencies. In particular, we expect to find that the spectrum of signals shows two types of statistical distributions: a normal or skewed distribution. The former describes a large number of harmonics around the fundamental pitch, and the latter describes stable patterns of frequencies in which the rules of the music are strictly followed during the composition and performance processes. Consequently, the importance of this study is that entropy values related to the spectrum of the signal must contain information that indicates statistical characteristics---statistical distributions that best describes the data---for understanding the virtuosity in music.

This paper is organized as followed: the Material section describes the experimental design and the database. In particular, we aim to present the criteria and sequential process for classifying guitarists into two groups. The Method section shows the procedure of using the spectral decomposition, the best fit analysis based on the Kolmogorv-Smirnov (KS) goodness of fit test~\cite{Massey1951}, and computing the Shannon entropy. The Results section displays the relationship between both groups of data. Finally, the Discussion section includes ideas to consider, and presents the conclusions.

\section*{Materials and Methods}
\subsection*{Materials}
\begin{figure}[ht]
\centering
\caption{Flowchart of the experimental design.}
\includegraphics[width=13.5 cm]{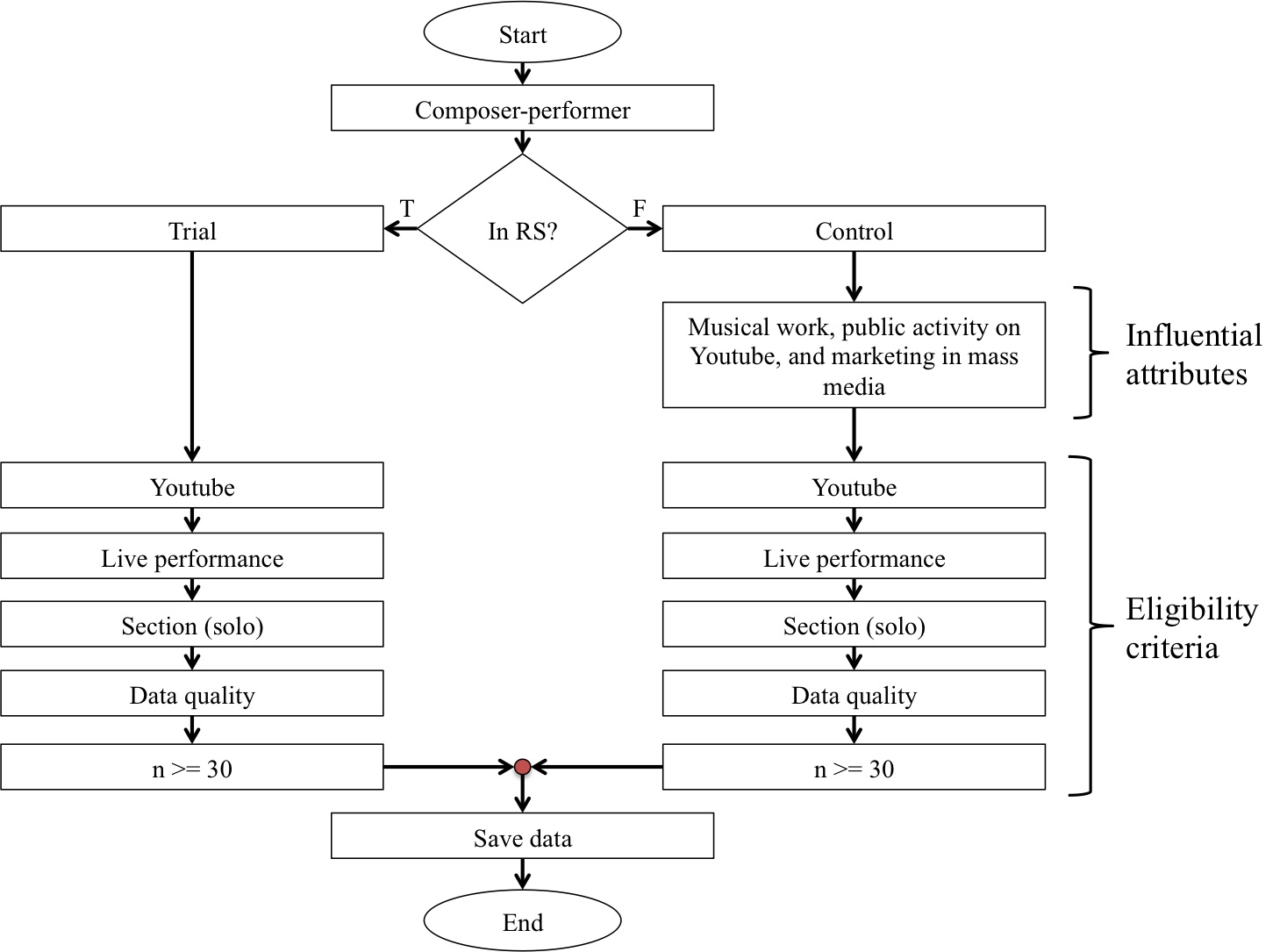}
\label{fig0}
\end{figure}

As we mention above, the concept of virtuosity has to be understood as a collective behavior of
guitarists who show the composition and performance abilities. However, the problem is to measure and identify the virtuosity based on the audio signals. Therefore, we propose an experimental design based on two sets of electric guitarists: the control and trial groups (Fig~\ref{fig0}). The former, named as ``all others,'' refers to a set of guitarists who are not consider ex ante as legends by the RS, and they are the baseline for determining the effectiveness of our study treatment. The latter, named as ``legends,'' refers to a set of guitarists who are considers ex ante as legends by the RS. Each group is based on 30 guitarists. The number of guitarists per group was selected based on the Central Limit Theorem in statistics that states $n>=30$, where $n$ is the number of observations in a sample~\cite{Ross2012,Ross2020}. The key assumption underlying this sample size is that the population distribution of the composer-performer guitarist from novice to expert is normal. We considered that this normal approximation is closely related to the age structure of the population variable, which describes the size of the population of a given age~\cite{owidagestructure}. Therefore, the population distribution of the composer-performer guitarists describes the size of the population of a given musical training, and this distribution can generally be characterized by its median and variance.

Next, the condition that allocates composer-performer guitarists into the two sets is related to the RS classification~\cite{RS2020}. We used this reference because it is one of the best non-scientific organized approximation to classify contemporary guitarists based on the opinion of their peers and people related to the music industry. 

Figure~\ref{fig0} presents the sequence of steps for selecting guitarists in both groups. The first step, represented by a rectangle, is to define our main set of guitarists to analyze. In our case, we are interested in musicians who show the composer-performer ability from novice to expert. Nowadays, we consider that this is
a common attribute in guitarists because the musical practice incorporates not only the theory or technical exercises, but also a deliberate, creative process of improving musical abilities. Compared to the Renaissance era, today people regularly can access to the musical practice. 

In the second step, we applied the condition related to the RS list, represented by the diamond. We have to recall that this list was generated based on a panel of top guitarists and people related to the music industry to rank their favorite guitarists. Then, this decision allocated guitarists into two groups. The trial group is related to those musicians previously categorized as legends by the RS, and the control group is associated with those musicians not included in the RS classification and possibly located into a subgroup of expert composer-performer guitarists. The selection process in the trial group was based on two steps.
The first was a deterministic selection that mapped guitarists from number one to ten. We applied this mapping selection because those guitarists are the top 10 of the RS list and the most famous, well-known, and respected guitar players of all times. For example, most of them are present in almost all of the guitarist's rankings~\cite{TG2020}. Even though we mapped those guitarists, not all of them were selected because they have to meet each of the eligibility criteria. For example, Chuck Berry is in position 7, but we did not find an audio signal that meets the criterion of data quality. In particular, we could not identify a clear sound of guitar solo. His signal was a mix of different instruments---guitars, drums, and basses---instead.
The second step, after passing the first ten guitarists in the RS list and be confident that we considered them, aims to select with the same probability the next guitarists until we stop the selection when $n >= 30$. 

On the other hand, the selection of guitarists in the control group was less constrained than in the trial. Because in the control group the selection of any expert composer-performer guitarist is possible, we aimed to select them with the same probability based on their current influential attributes: musical work, public activity in youtube, and marketing in mass media. We use the \href{https://www.google.com/search/howsearchworks/}{Google search} engine that is based on the \href{https://networkx.org/documentation/stable/reference/algorithms/generated/networkx.algorithms.link_analysis.pagerank_alg.pagerank.html#:~:text=PageRank\%20computes\%20a\%20ranking\%20of,algorithm\%20to\%20rank\%20web\%20pages.&text=A\%20NetworkX\%20graph.,edges\%20for\%20each\%20undirected\%20edge}{PageRank} algorithm for identifying the influential attributes. For example, base on the guitarists' name, we can recognize such attributes based on the searching results. Inevitably, the process to select those guitarists shows a possible selection bias related to popularity and demographics. To alleviate this potential bias, we aimed to consider as many guitarists as we can easily identify by the Google searching engine. For example, John Petrucci, a founding member of the band Dream Theater, continuously produces and publishes videos in youtube that promote his personal activities as well as related collaborations, for example the promotion of his guitar collection with the Ernie Ball Music Man~\cite{MM2022}. Another example is Gustavo Cerati, former lead singer and guitarist of the Argentine rock band Soda Stereo. Even though he passed away in 2014, his music is continuously updated in his youtube channel. Both are well-known and respected guitarists, but they are outside of the RS list. Therefore, we selected different guitarists until we reach the criterion of $n >= 30$.

Next, for each group, we used the well-known video streaming~\href{https://www.youtube.com/}{youtube} as our main data source. By using this data source, we can search for different types of videos.
We searched for a video or audio related to live performances. Compared with video or audio records in a studio, which is a controlled environment for recording audio signals, we selected live performances because they are an uncontrolled environment in which the musician expresses its abilities and experience in a particular context and time period. For example, the work of Swarbrick et al.~\cite{Swarbricketal2019} suggested that ``\dots live music engages listeners to a greater extent than pre-recorded music and that a pre-existing admiration for the performers also leads to higher engagement.'' Therefore, in live performances not only the listener experiences an affiliative social engagement, but also the musician experiences increasing feelings of involvement when playing in a unique and not predetermined way.

Moreover, another filter that we used after identifying the video or audio related to a live performance was to select an instrumental section or ``solo'', between 30 sec. and 3 min. approximately, of a piece of music. This section contains not only the techniques but also the improvisation related to the musical expression. Then, it is possible to select the complete section or a subsection of it, if and only if, it presents these characteristics. Subsequently, after considering the composer-performer main set, the youtube as data source, the live performance, and the ``solo'' section, we examined the quality of data. In particular, we searched for the best possible, clearest rendition of audio. At the end of this process, some audio signals were cleaned to improve their signal by a noise reduction function (see the Supplementary Information, Table~\ref{tableArpStat}, \ref{tableTrial}, \ref{tableControl})~\cite{audacity2022}. 

The final step in this sequence of processes is to save the selected data. We recorded the audio signal of each selection using~\href{https://www.audacityteam.org/}{Audacity} in a Macbook pro computer.
We selected the waveform audio file (WAV) and used the mono audio as the sound output when recording the signal. Files were saved as 16 bit PCM. We used this format because it is a standard, simple, accurate, and uncompressed computer audio file. However, there are other options to record sound waves, for example Audio Interchange File Format (AIFF), Raw Audio Format (RAW), or Broadcast Wave Format (BWF), among others. Therefore, the data we used as the audio source is available in youtube (see the reference of videos in the Supporting information section and our youtube list: \href{https://youtube.com/playlist?list=PLesSAX_bg1DX4U1gPR_bKTbM3QMkMyQQ-}{Virtuosity (trial group)} and \href{https://youtube.com/playlist?list=PLesSAX_bg1DVWSA5KkJrn9OcVo2S5Yhcn}{Virtuosity (control group)}).

In order to further explain and present the reproducibility of this experiment, we exemplify it by selecting one musician per group. We start by considering the composer-performer abilities. Then, we use the RS list and take the first guitarist in the list. Next, we check if this guitarist has a video in youtube that fulfills the eligibility criteria: a live performance, a ``solo,'' and a good quality of video or audio. Keeping in mind these criteria, we selected the audio signal and allocated the guitarist into the trial group. As a result, our first trial guitarist was Jimi Hendrix. In the case of the control group, we first must consider an expert guitarist that has the influential attributes. Keeping in mind these, we must also check for the eligibility criteria. We considered in first place Alejandro Marcovich, an Argentinian-Mexican guitarist and founding member of the Mexican band Caifanes. We selected him because he has the influential attributes and has a video or audio that meets with the eligibility criteria. After searching and testing these criteria, we had to decide whether to include or not his audio signal into the control group. In the case of Marcovich, we included his data in the control group. Therefore, this process is repeated until we meet the criterion of $n >= 30$ in both groups (see the final selection of both groups in the Supplementary Information, Table~\ref{tableTrial} and~\ref{tableControl}).

\begin{figure}[ht]
\centering
\caption{Arpeggio.(a) Waveform, (b) Spectrum in a linear scale, and (c) log-log scale. Resource: \href{https://freesound.org/people/Lucks86/sounds/246268/}{https://freesound.org/people/Lucks86/sounds/246268/}}.
\includegraphics[width=13.5 cm]{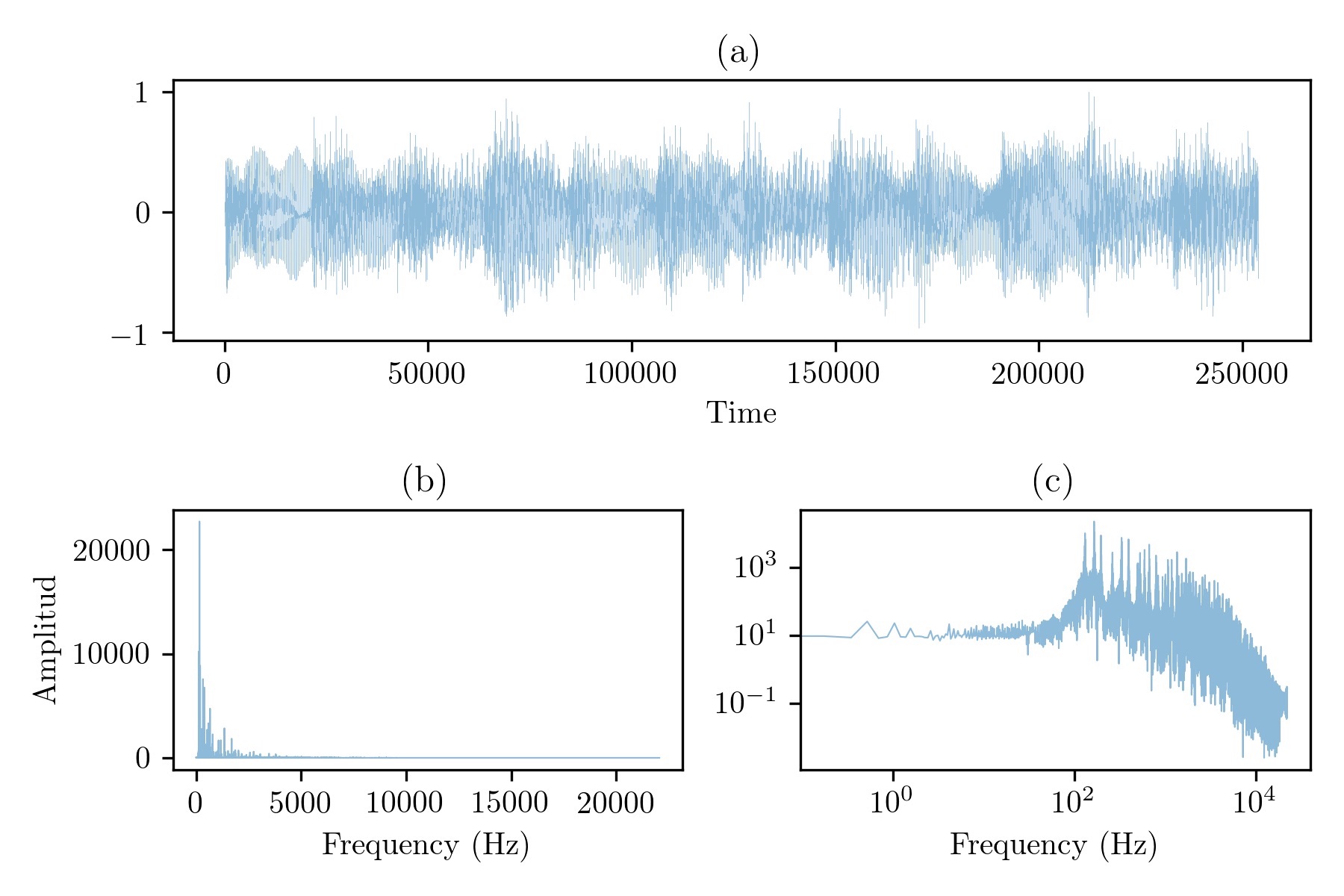}
\label{fig1}
\end{figure}

\begin{figure}[ht]
\centering
\caption{Gaussian noise.(a) Waveform, (b) Spectrum in a linear scale, and (c) log-log scale. Resource: \href{https://freesound.org/people/Jace/sounds/35291/}{https://freesound.org/people/Jace/sounds/35291/}.}
\includegraphics[width=13.5 cm]{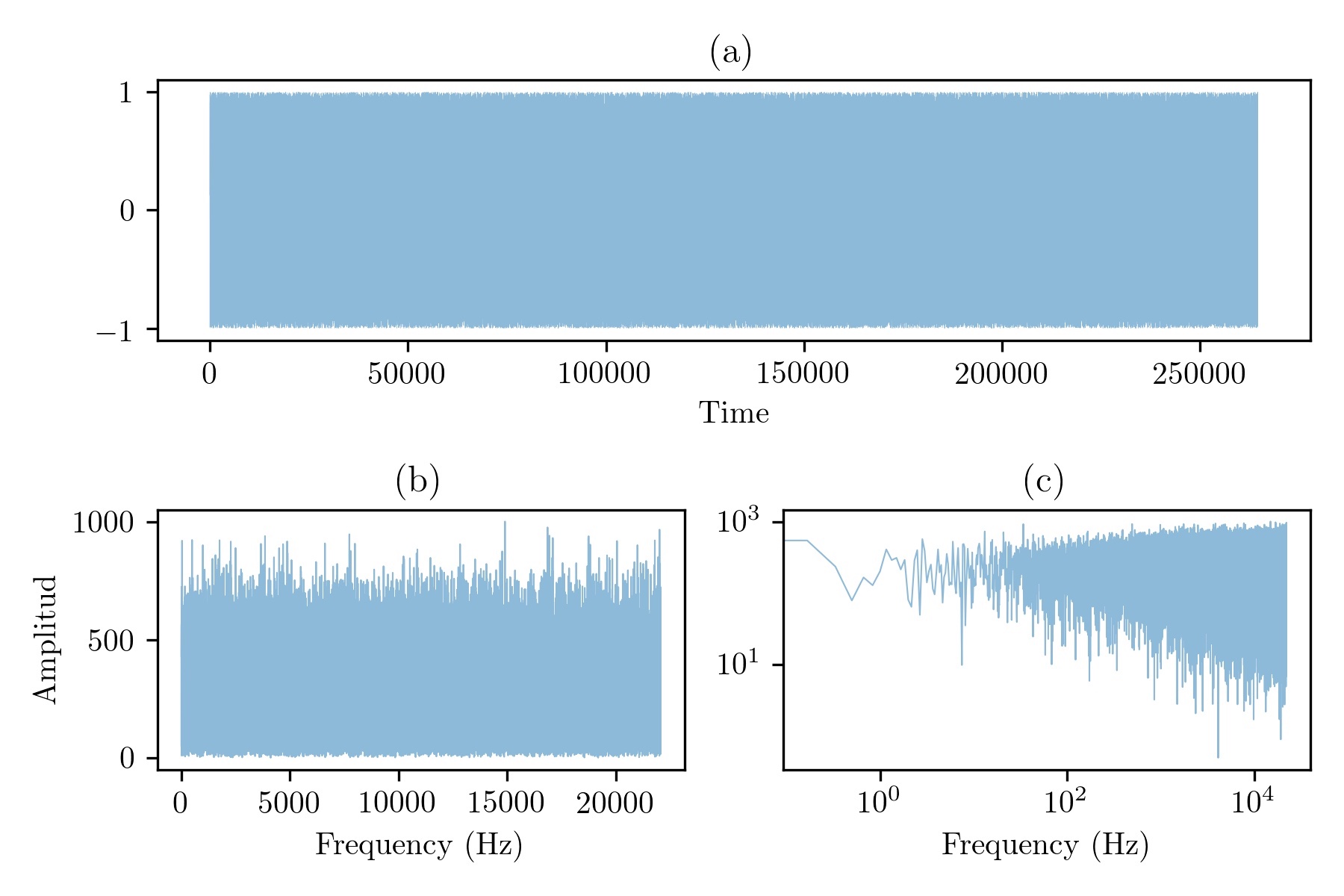}
\label{fig2}
\end{figure}

In addition, to compare our database with well-known signals and to validate part of our analysis (see the Methods section), we selected the arpeggio and the Gaussian noise signals. They were selected from the \href{https://freesound.org/}{freesound.org} database. The signals show the following behaviors (Fig~\ref{fig1}, and Fig~\ref{fig2}).

Fig~\ref{fig1} shows the attributes of the arpeggio signal. Subfigure (a) and (b) are the images of the waveform and its spectrum. Subfigure (c) shows the log-log plot that clearly describes different regions of the spectrum---a flat section and a negative trend.

Fig~\ref{fig2} displays the attributes of the Gaussian noise signal. As we described in Fig~\ref{fig1}, subfigure (a) and (b) are images of the waveform and the spectral decomposition. They approximately show equal proportions of each frequency that are present in the signal. Subfigure (c) shows a stable behavior of the spectrum. 

On the other hand, in the case of code libraries, we used different Python libraries to retrieve, analyze, and plot the data and results. In particular, we used the \href{https://numpy.org/}{Numpy}, \href{https://matplotlib.org/}{Matplotlib}, \href{https://www.scipy.org/}{Scipy}, and \href{https://pandas.pydata.org/}{Pandas}. Furthermore, we used some part of the code provided by Downey~\cite{Downey2016}. The database and the code are available in our Open Science Framework (OSF) for the reproduction of our findings: \href{https://osf.io/9pn58/?view_only=3d3160a3a9bc4892abe5335d90b75a66}{Complex systems and music}.

Finally, it is important to mention that we did not consider the music notation as an input for our analysis because it is a different part of the music. The music notation is the language of the human to understand and communicate sounds. Therefore, it is the human interpretation of the pitch, and it deserves a different approximation and data analysis.

\subsection*{Methods}
We consider a music signal that represents the output of the guitar player. This type of signal contains basic data that we can use to identify and measure attributes related to the notion of virtuosity. Therefore, we propose a method based on three sequential processes applied to the signal of both groups. The first is to apply the spectral decomposition based on the FFT~\cite{Cooleyetal1965, Press2007}, the second is to identify the statistical distribution that best describes the spectrum based on the KS test~\cite{Massey1951}, and the third is to compute a diversity measure based on the Shannon entropy~\cite{Shannon1948} and compare it between and within groups. It is important to note that this last comparison aims to complement our exploratory data analysis. It is not intended as a paired test study as it is routinely done in randomized controlled trial (RCT) studies.

The spectral decomposition associated with the FFT is a transformation that decomposes the signal into a series of frequency components that may or may not coincide with the true frequencies in the signal. This simplification is a very suitable approximation for identifying each pitch with its frequency in Hz and its magnitude. The FFT is based on the following formulation~\cite{SciPy2021}:

\begin{eqnarray}
\label{eq:1}
	y[k] = \sum_{n=0}^{N-1} e^{-2\pi j \frac{kn}{N}} x[n]
\end{eqnarray}
in which $y[k]$ is the frequency component of a sequence of the signal $x[n]$ from $n$ to $N-1$. In our case, the richness of the timbre detected using the FFT can indicate levels of virtuosity in music associated with different sizes of harmonic sets.

Next, based on this data, we analyzed the statistical distributions that best describe it. We compared the magnitude values with a set of 12 continuous statistical distributions~\cite{LugoAlatriste2020}. These distributions are the following: 1) normal, 2) lognormal, 3) gamma, 4), exponential, 5) Pareto, 6) Gilbrat, 7) power law, 8) exponentiated Weibull, 9) Weibull minimum, 10) Weibull maximum, 11) beta, and 12) uniform. We applied the KS goodness of fit test that compares each signal with each of these statistical distributions. Then, for every signal, we identified the best and second best fit (see the Supplementary Information, Table~\ref{tableBFArpStat}, \ref{tableBFTrial} and \ref{tableBFControl}). This process determines the statistical distribution that best describes the data.

Finally, we used the Shannon entropy measure to identify levels of diversity~\cite{Shannon1948}. In our context, we used the concept of diversity to identify how close or far are the signals to the arpeggio and Gaussian noise signals mentioned previously (Fig~\ref{fig1} and~\ref{fig2}). We compute this measure following the next formulation:

\begin{eqnarray}
\label{eq:2}
	H = - \sum_{i=1}^{n} p_{i} log_{} p_{i}
\end{eqnarray}
in which $H$ is the entropy, $p_{i}$ is the relative frequency of $i$ associated with the \emph{n}-items of given intervals $n$. Compared with the original formulation of Shannon~\cite{Shannon1948} who used $log_{2}$, we use $log_{10}$ because we are interested on making evident the differences between the frequency components. This change of scale---a monotonic transformation that preserves the order of the numbers---help us to clearly identify the entropy values of music information. In particular, the range of values decreases from $log_{2}$ to $log_{10}$. However, the interpretation in both scales are the same. Then, decreasing values of $H$ indicate less diversity close to the arpeggio value. On the other hand, larger values of $H$ denote higher diversity close to the Gaussian noise value. Moreover, we used the \href{https://docs.scipy.org/doc/scipy/reference/generated/scipy.stats.mannwhitneyu.html}{Mann-Whitney U rank test} to compare entropy values between groups~\cite{MannWhitney1947,FayProschan2010}.

Therefore, we are going to use the output of the Eq~\ref{eq:1} as an input for computing the KS test and the diversity (Eq.~\ref{eq:2}). In the next section, we present our results that compare both groups based on the method above  (see the code in the OSF project \href{https://osf.io/ke9uj/?view_only=3d3160a3a9bc4892abe5335d90b75a66}{Complex systems and music}).

\section*{Results}
Based on the two sets of guitarists, we analyze their signals and show our findings regarding the search for  relationships between the two groups, and for trends that exist in the virtuosity of the musical performances and the related entropy of the associated musical information. In this section, we present the best fit statistical distribution and the entropy findings. In addition, to illustrate the results at a finer scale, we present two cases per group of guitar players.

\begin{figure}[ht]
\centering
\caption{Number of guitar players in the trial and the control group associated with the best fit statistical distribution. See the Supplementary Information, Table~\ref{tableBFTrial} and \ref{tableBFControl}, for details of the estimated parameters and the KS goodness of fit test values. See the Supplementary Information, Table~\ref{pdfs} for the probability density functions (PDF) associated with the statistical distributions.}
\includegraphics[width=13.5 cm]{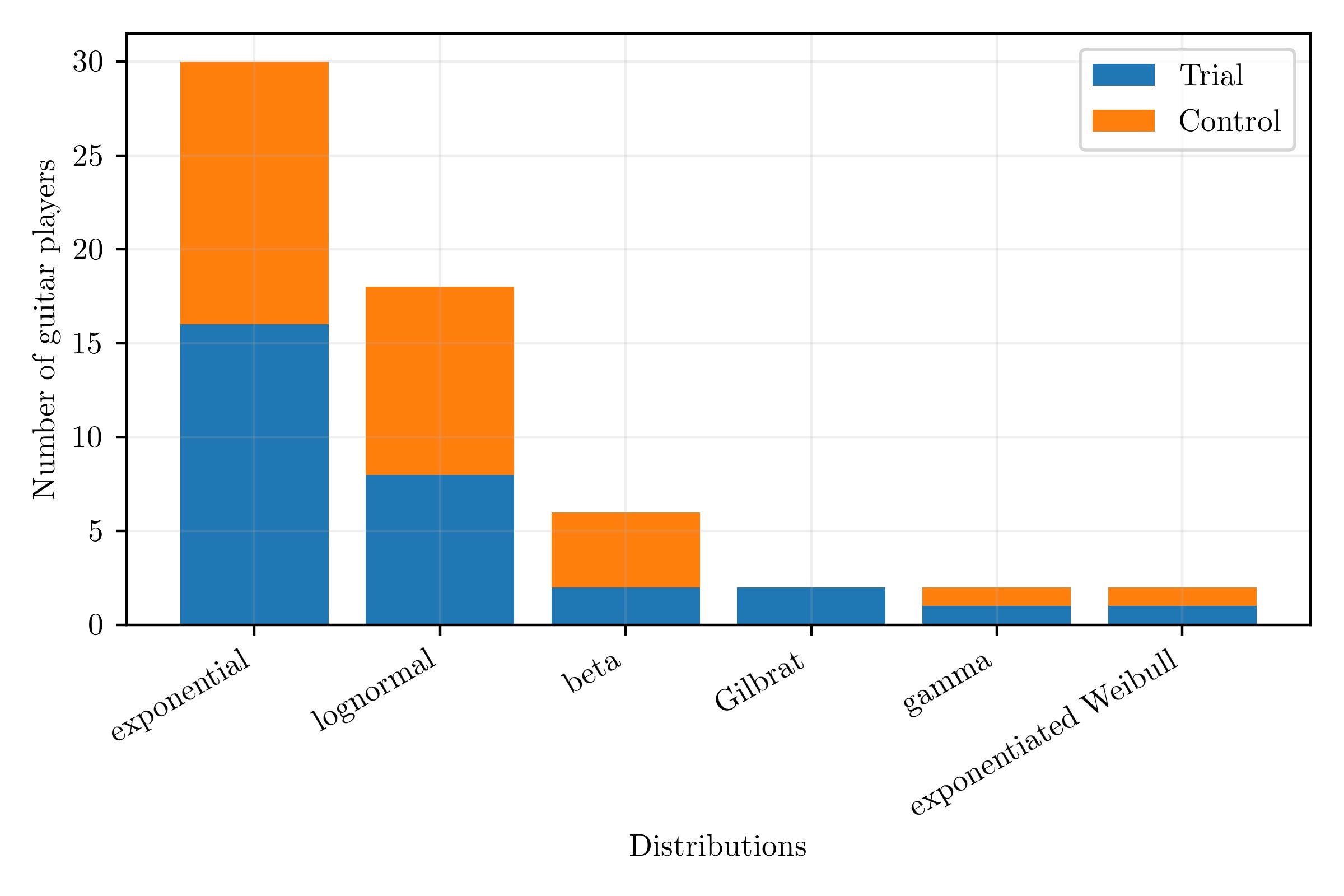}
\label{fig3}
\end{figure}

The first result is the frequency of the best fit statistical distributions found in our analysis (Fig~\ref{fig3}). In particular, Fig~\ref{fig3} shows the number of guitar players per group associated with such distributions. 

Fig~\ref{fig3} reveals that the exponential is the most frequent distribution related to the trial and control groups. In particular, the trial and control groups show 16 and 14 guitarists, respectively. The second most frequent distribution is the lognormal showing 8 players in the trial and 10 in the control. The third is the beta showing 2 and 4 guitarists in the trial and control group, respectively. The fourth is the Gilbrat associated with the trial group. The rest of the distributions, gamma and exponentiated Weibull, show small and similar proportions between the groups. Therefore, the trial and control groups show similar statistical distributions in proportion to the number of guitarists.  
 
\begin{figure}[ht]
\centering
\caption{Diversity according to the trial and control groups of guitar players and their best fit statistical distribution. Subfig (a) and (b) show the guitarists and their best fit of the trial and control groups from low to high values. The arpeggio and the Gaussian noise are the lowest and highest values in diversity respectively. Subfig (c) shows the comparison of the trial and control groups by pair of guitarists without the arpeggio and the Gaussian noise value.}
\includegraphics[width=13.5 cm]{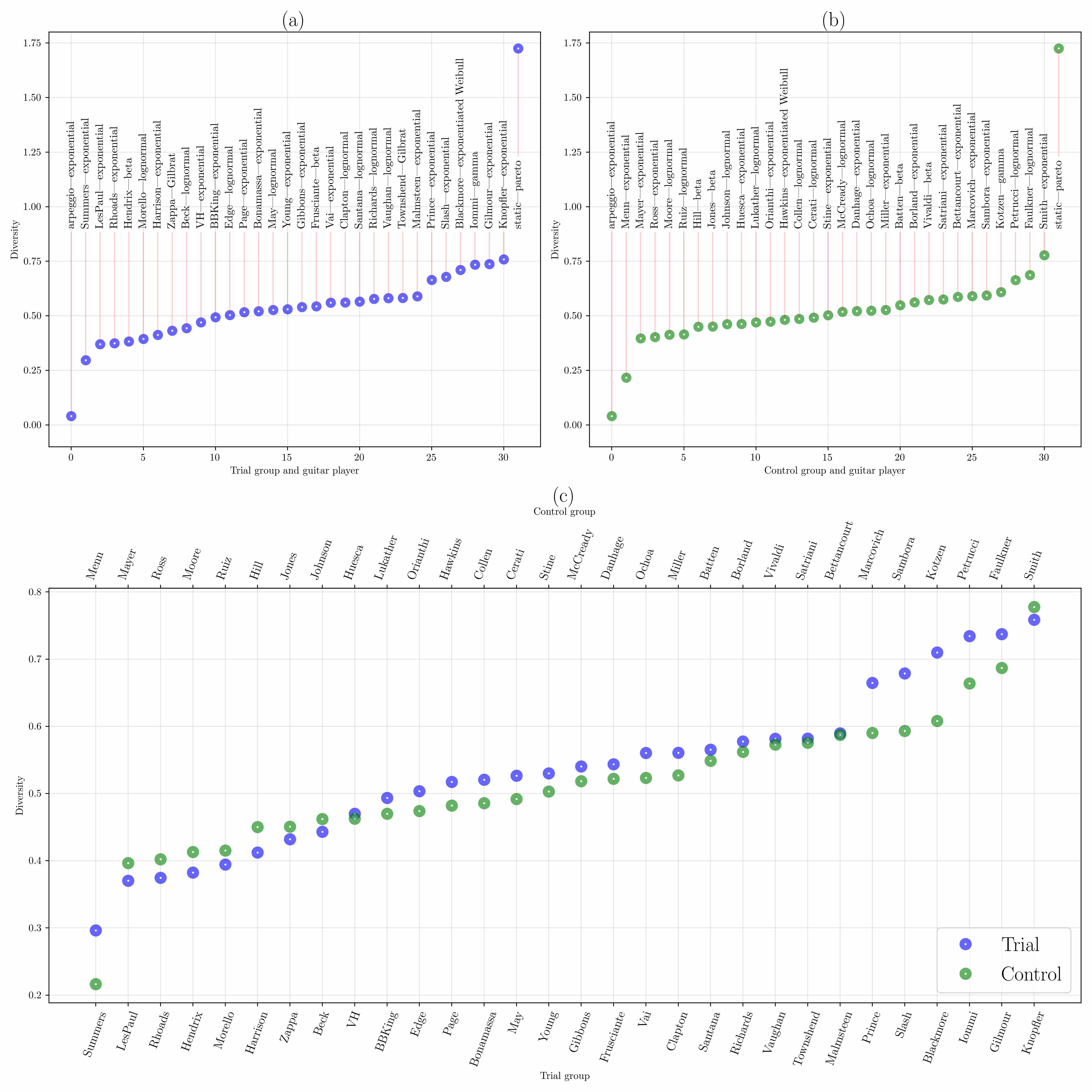}
\label{fig4}
\end{figure}

To complement Fig~\ref{fig3}, we display the entropy based on the diversity value of the two groups of guitar players and their best fit statistical distribution (Fig~\ref{fig4}(a), Fig~\ref{fig4} (b), and Fig~\ref{fig4} (c)). 

Fig~\ref{fig4}(a) and (b) show a slight rise in the number of diversity values of the trial and control groups. Fig~\ref{fig4}(a) displays the value of Summers slightly below the rest of the values, and from Prince to Knopfler, the diversity is rapidly increasing in values. Fig~\ref{fig4}(b) shows similar behavior compared to Fig~\ref{fig4}(a) for the Menn, Petrucci, Faulkner, and Smith, respectively. Finally, we see a slight difference between values of guitarists of both groups and the arpeggio and Gaussian noise values.

Fig~\ref{fig4}(c) shows the closeness of both groups in terms of their entropy. Both display a similar trend and close values of entropy. These values identify a range between 0.2 and 0.8 in which the entropy of both groups are located. However, we can see a different behavior when comparing a pair of guitarists. In particular, some sections of points shows a crosswise pattern in which values of one group are below and above the other group's values. This particular pattern suggests that groups are not different in the distribution underlying them. Therefore, Table~\ref{tab1a} and~\ref{tab2a} confirms the similarity.

\begin{table}[ht]
\centering
\begin{tabular}{|c|c|c|c|c|}
\hline
\textbf{Group}	& \textbf{Best fit}	& \textbf{Parameters}     & \textbf{KS test} & \textbf{First moments}\\
\hline
$Trial$	& gamma & (122.44451502094967, & (0.12914869508242166, & (\textbf{0.5314125754},\\
	&  & -0.7485414426808097, & 0.6770027753472341) & 0.53490483,\\
	&  & 0.010481860092354429) &  & 0.0134529,\\
	&  &  &  & 0.18074252, \\
	&  &  &  &  0.04900179)\\
\hline

$Control$ & gamma & (900.1898213776649, & (0.12909839636580123, & (\textbf{0.5129645989},\\
	&  & -2.5676631398085332, & 0.6775534048290307) & 0.51410568,\\
	&  & 0.0034234655242896435) &  & 0.01055033,\\
	&  &  &  & 0.06665964, \\
	&  &  &  & 0.00666526) \\
\hline
\end{tabular}
\caption{\label{tab1a}Entropy, best fit, KS test, and first moments. Group is related to trial and control data, the best fit shows the name of the statistical distribution, parameters show the estimated parameters of the best fit distribution, KS test displays the KS goodness-of-fit test, and first moments show the statistics of median, mean, variance, skewness, kurtosis. Bold values in the First moments column indicates the Median (see the Supplementary Information, Table~\ref{bestFitMWUtFull} for the first and second best fit results.}
\end{table}

\begin{table}[ht]
\centering
\begin{tabular}{|c|c|}
\hline
\textbf{$U_{Trial}$}	& \textbf{p-value}\\
\hline
488.0	& 0.5792942105819812 \\
\hline
\end{tabular}
\caption{\label{tab2a}Mann-Whitney U test. The Mann-Whitney U two-sided test is based on the following null hypotheses: $H_{0}$:  The two data are equal. We decided a confidence level of 95\% to reject the null hypothesis in favor of the alternative that distributions are different.}
\end{table}

Table~\ref{tab1a} shows the similar name of the best fit distribution in both groups. However, the estimated parameters reports small differences of values. In addition, if we observe the result of the Mann-Whitney U test (table~\ref{tab2a}), we confirm that the two values of groups are similar. We confirm $H_{0}$ that the two samples have the same shape. 
Therefore, entropy based on the diversity values of both groups are close to each other.

\begin{figure}[ht]
\centering
\caption{Spectrum of a pair of guitarists with low values of entropy. Spectrum on a linear scale. For the purpose of clarification, these values are the result of using Eq~\ref{eq:1}.}
\includegraphics[width=13.5 cm]{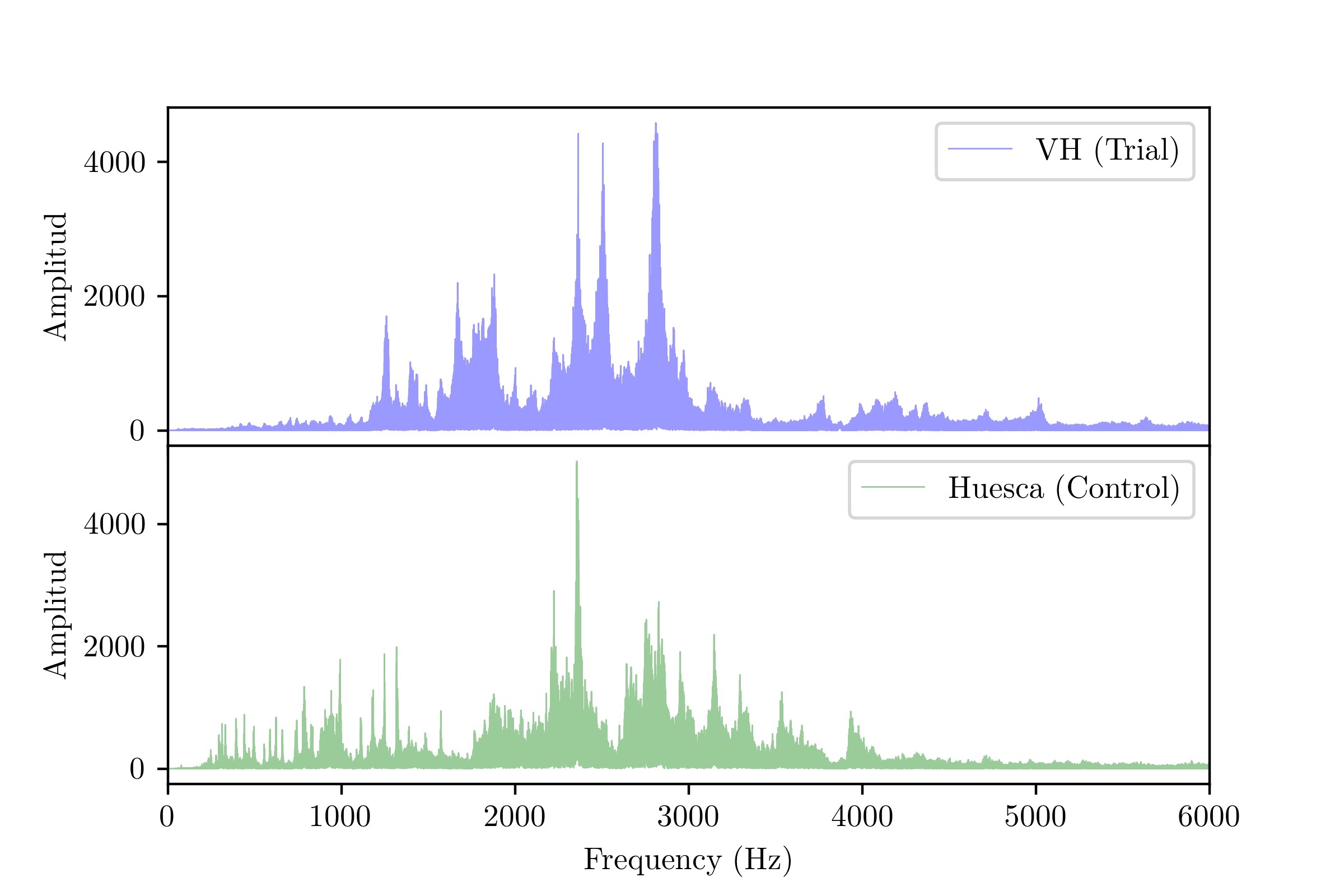}
\label{fig5}
\end{figure}

\begin{figure}[ht]
\centering
\caption{Relative frequency distributions of guitarists with low values of entropy.
Data is related to the frequency components of the spectrum (Fig~\ref{fig5}). Bins are related to the system called the~\href{https://pages.mtu.edu/~suits/notefreqs.html}{12-Tone Equal Temperament} that shows the number of oscillation per second (Hz) and its associated note. The note C with a number represents higher and lower registers in the audible range.}
\includegraphics[width=13.5 cm]{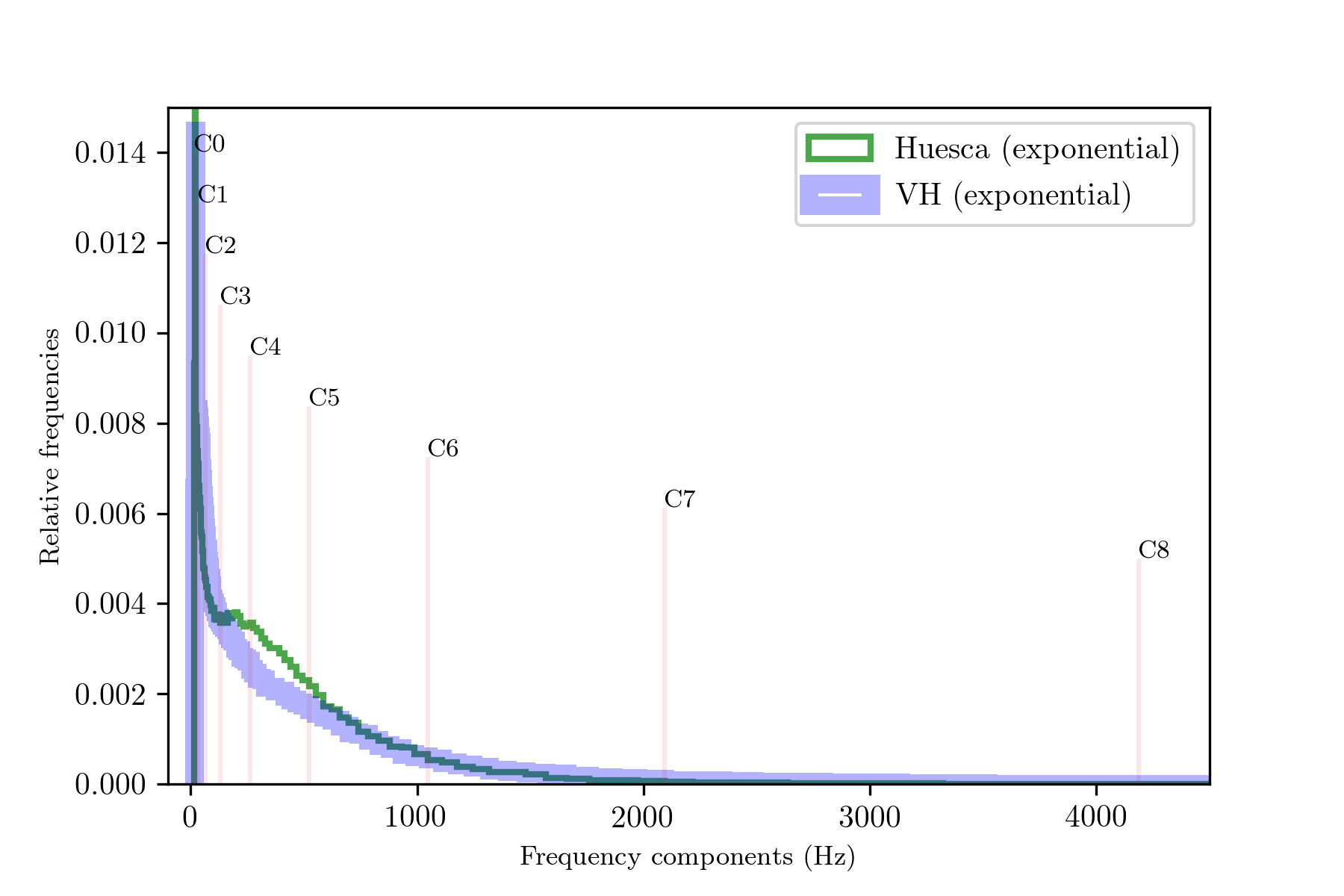}
\label{fig8}
\end{figure}

\begin{figure}[ht]
\centering
\caption{Spectrum of a pair of guitarists with high values of entropy. Spectrum on a linear scale.}
\includegraphics[width=13.5 cm]{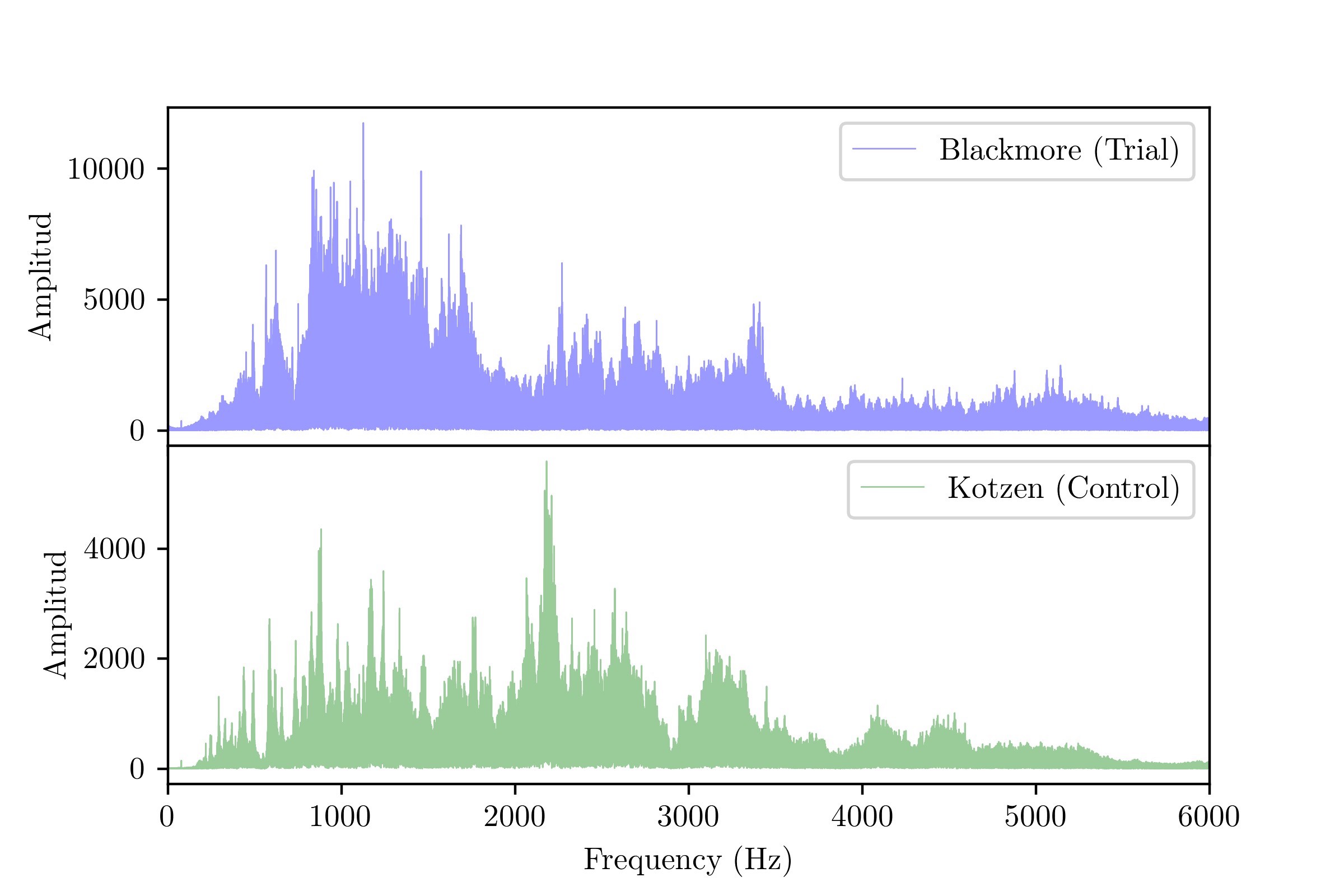}
\label{fig6}
\end{figure}

\begin{figure}[ht]
\centering
\caption{Relative frequency distributions of guitarists with high values of entropy.  
Data is related to the frequency components of the spectrum (Fig~\ref{fig6}). Bins are related to the system called the~\href{https://pages.mtu.edu/~suits/notefreqs.html}{12-Tone Equal Temperament} that shows the number of oscillation per second (Hz) and its associated note. The note C with a number represents higher and lower registers in the audible range.}
\includegraphics[width=13.5 cm]{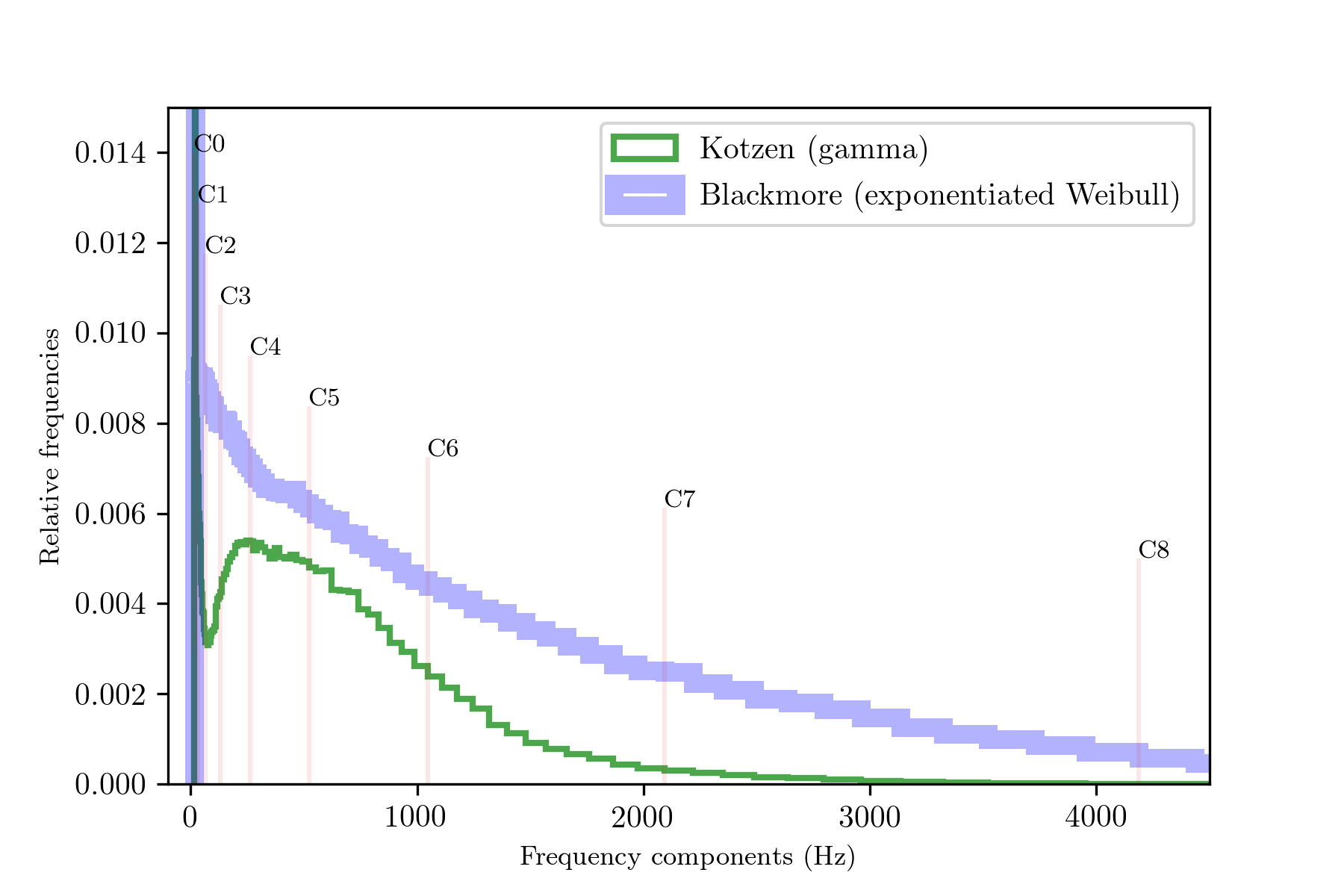}
\label{fig9}
\end{figure}

Having identified similarities between groups, we will now show particular cases to exemplify these findings. Based on Fig~\ref{fig4}(c), we selected a pair of guitarists of both groups with low values of entropy (Fig~\ref{fig5} and Fig~\ref{fig8}) and a pair of guitarists with high values (Fig~\ref{fig6} and Fig~\ref{fig9}). In particular, the former pair of guitarists are VH and Huesca due to their similar diversity values. The latter is a pair of guitarists with a separated entropy values, Blackmore and Kotzen. 

Fig~\ref{fig5} shows the spectrum of VH and Huesca signals related to a pair of low diversity values, 0.4699 and 0.4625 respectively. The distribution that best describes both frequencies in the spectral is the exponential. An important characteristic of this result is the clear identification of dominant frequencies and their concentration. In particular, the spectrum of VH displays three dominant frequencies, higher than 2500 of Amplitude and between 2400 and 3000 Hz. On the other hand, the spectrum of Huesca displays a clear dominant frequency that is similar to the one of the VH dominant frequencies. Therefore, lower values of diversity are associated with homogeneous and concentrated frequencies in the spectrum of the signal. Some similarities between groups and guitarists are more evident.

Fig~\ref{fig8} shows the statistical distributions of frequency components based on the VH and Huesca signals. Both data show skewed distributions and similar shapes. It can be seen from C3 to C5 that the distributions are different from each other. This result indicates that Huesca more frequently used notes of such a range of register than VH. Therefore, the Huesca's case exemplifies particular ups and downs in the distribution pattern related to the more frequent number of notes played by the guitarist in a particular range of register. On the other hand, the case of VH exemplifies the common pattern of a gradual decrease in the number of frequency components in which higher and lower registers are more frequently played, i.e., the presence of extreme events.   

Fig~\ref{fig6} reveals the spectrum of Blackmore and Kotzen signals. They are related to diversity values of 0.7100 and 0.6081 and the best fit distributions of exponentiated Weibull and gamma, respectively. It is apparent from this result that the appearance of these spectrums are different compared to the spectrums in Fig~\ref{fig5}. In particular, the behavior of the spectrum reveals variable and dispersed frequencies values. The case of Blackmore presents these characteristics. The dominant frequencies and their harmonics are not evident. On the other hand, the case of Kotzen shows a less diffused behavior in frequencies where we can identify one dominant frequency higher than 4000 of amplitude. Therefore, higher values of diversity are associated with variable and dispersed frequencies in the spectrum of the signal. Some differences between groups and guitarists can be more evident.

Fig~\ref{fig9} shows the statistical distributions based on the Blackmore and Kotzen signals. As it can be seen from this data, the differences between both distributions are more clear than in Fig~\ref{fig8}. The relative frequencies of the number of frequency components in the Blackmore signal are higher than in Kotzen. Moreover, from C3 to C4, the Kotzen distribution shows a marked increase in the relative frequencies of frequency components. Form C4 to C6, the relative frequencies show a gradual decrease in the number of frequency components. Then, in the range of C3 to C5 the Kotzen data displays ups and downs in the distribution pattern. Therefore, different best fit statistical distributions can be associated with higher values of diversity.         

In summary, we showed in this section that there is a range of entropy values where the two groups of guitarists are located. These groups share the same type of statistical distribution, even though they show small differences of values. Those values are related to their spectrum: homogeneous-concentrated and heterogeneous-dispersed frequencies.

\section*{Discussion}
The present study was designed to explore the possibility of measuring and detecting the virtuosity in a collective set of audio signals generated by composer-performer guitarists. Comparing the trial and control groups, we found similar values of entropy located in a particular range and best fit statistical distributions.

Firstly, considering entropy values associated with the diversity, our data analysis identified a range in which such values are located. This finding suggested that musicians, who have the composer-performer abilities, might be located within this range. Then, the virtuosity in music is possibly expressed as a quantitative attribute within a range of frequency components of the audio signal. Therefore, our findings suggested that the virtuosity in music of a collective audio signal can be perceived by entropy values. For example, any musician with those abilities can compute the entropy value and identify the location inside the virtuosity range. However, each time when the guitarist played and recorded the sound as audio signal, the value of entropy may vary because it depends on multiple factors, for example the environment and the musician mood.

About the best fit statistical distribution, our study confirmed a non-significant difference between the trial and control group. In spite of the opinion of a collective and experienced group of guitarists for classifying their colleagues, these results suggested that the audio signal underlies the basic music information for better understanding the playing of guitarists. From novice to expert, composer-performer musicians show different levels of entropy, and consequently a range of virtuosity levels. In addition, we identified the type of statistical distribution of each frequency component of audio signal in which the most frequent distributions are the exponential and lognormal. Both distributions are different from each other, but both describe skewed distributions. The lognormal represents a special case of the normal distribution. On the other hand, the exponential distribution shows asymptotic relationships with the Weibull and Pareto distribution. Therefore, our collective analysis of audio signals that best describe their spectrum reinforces the idea that the virtuosity in music might be identified by analyzing the entropy values.  

We did not address in this study the generalization of our results to other instruments, music styles, and composer-performer musicians. It is not trivial to extent our analysis for justifying the entropy as a general measure. We should collect a big database per instrument, music style, and musician, and analyze each of them. Moreover, we should consider different types of machine learning methods for validating and testing data. Therefore, further work is needed to fully understand the implications of the relationship between the virtuosity and entropy. For the short-term work, we plan to study classical instruments and musicians. For example, the instrument of the bass viola du gamba and the Marin Marais and Bach musicians, played by Jordi Savall. For the long-term work, we plan to generate a big database of audio signal for embracing the transparency, openness, and reproducibility of science and music. 

\section*{Conclusions}
Guitar players who compose and perform extraordinary pieces of music are rare. However, nowadays we have the technology and the knowledge to analyze audio signals and decipher other dimensions of the music. In particular, this type of signal encapsulates the concept of virtuosity, which can be identified and computed for any musician to better understand the personal process of composition and performance. 

In addition, the creation process, using an electric guitar, most of the time starts with a simple idea translated into chord progressions and phrasing arpeggios. Subsequently, the dynamics of composing and performing generate multiple outcomes in which the creativity arises from simple ideas. In other words, the physical and cognitive abilities synchronize and generate a natural or organic performance that transcends all previous processes. Therefore, by associating the virtuosity with values of entropy, we are able to observe, to hear, and measure the last part of this creative process.

\section*{Additional information}
Data available in a publicly accessible repository. The data presented in this study are openly available in [Open Science Framework] at [\href{https://osf.io/9pn58/?view_only=3d3160a3a9bc4892abe5335d90b75a66}{Complex systems and music}]. DOI's will be available after the manuscript is accepted for publication.

\section*{Supplementary Information}
\paragraph{See the supplementary information document.}
\label{SI}

\newpage
\bibliography{mybibfile}

\section*{Acknowledgements}

We would like to thank Gustavo Martinez-Mekler for his comments.

\section*{Author contributions statement}

Conceptualization, I.L and M.A-C; Formal analysis, I.L and M.A-C; Investigation, I.L.; Methodology, I.L and M.A-C; Validation, I.L and M.A-C; Visualization, I.L; Writing--original draft, I.L and M.A-C; Writing--review \& editing, I.L and M.A-C.

\section*{Supplementary Information}

\subsection*{Audio}

\paragraph{S1 Audio.}
\label{S1_Audio}
{\bf freesound.}  guitar arpeggio C.wav. [Audio]. freesound. Retrieved July 7, 2021, from \\ \href{https://freesound.org/people/Lucks86/sounds/246268/}{https://freesound.org/people/Lucks86/sounds/246268/}

\paragraph{S2 Audio.}
\label{S2_Audio}
{\bf freesound.}  continuous static.wav. [Audio]. freesound. Retrieved July 7, 2021, from \\ \href{https://freesound.org/people/Jace/sounds/35291/}{https://freesound.org/people/Jace/sounds/35291/}

%%%%%%Video
\subsection*{Video}

\paragraph{S1 Video.}
\label{S1_Video}
{\bf Guitarist.}  Jennifer Batten: Rig Tour. [Video].Youtube. Retrieved July 7, 2021, from \href{https://youtu.be/6oRn0VBYYiE}{https://youtu.be/6oRn0VBYYiE}.

\paragraph{S2 Video.}
\label{S2_Video}
{\bf B.B. King.} B.B. King - Sweet Little Angel (Live). [Video].Youtube.
Retrieved July 7, 2021, from \\ \href{https://youtu.be/dNr_eIgP0tI}{https://youtu.be/dNr\_eIgP0tI}.

\paragraph{S3 Video.}
\label{S3_Video}
{\bf feedbackbro.} Jeff Beck - Where Were You - (Live at Ronnie Scott's). [Video].Youtube.
Retrieved August 22, 2021, from \href{https://youtu.be/howz7gVecjE}{https://youtu.be/howz7gVecjE}.

\paragraph{S4 Video.}
\label{S4_Video}
{\bf tzgon.} YG 2014.06 Nuno Bettencourt. [Video].Youtube. Retrieved July 7, 2021, from \href{https://youtu.be/qAqEhC2u9_w}{https://youtu.be/qAqEhC2u9\_w}.

\paragraph{S5 Video.}
\label{S5_Video}
{\bf LiveOnMp3.} Ritchie Blackmore Guitar God. [Video].Youtube.
Retrieved July 7, 2021, from \\ \href{https://youtu.be/BIOeFTBQPHk}{https://youtu.be/BIOeFTBQPHk}.

\paragraph{S6 Video.}
\label{S6_Video}
{\bf Lucas JRL.} Joe Bonamassa Soloing. [Video].Youtube. Retrieved July 7, 2021, from
\href{https://youtu.be/hSKNiA5aTp8}{https://youtu.be/hSKNiA5aTp8}.

\paragraph{S7 Video.}
\label{S7_Video}
{\bf JackCarver10.} Wes Borland - Ibanez demonstration (without stupid My Way intersections). [Video].Youtube. Retrieved July 7, 2021, from \href{https://youtu.be/_vNUWWasENw}{https://youtu.be/\_vNUWWasENw}

\paragraph{S8 Video.}
\label{S8_Video}
{\bf Germán Ospina.} En La Ciudad De La Furia (Mtv Unplugged) - Soda Stereo Ft. Andrea Echeverry. [Video].Youtube. Retrieved July 7, 2021, from
\href{https://youtu.be/IlSWxzxl5cg}{https://youtu.be/IlSWxzxl5cg}.

\paragraph{S9 Video.}
\label{S9_Video}
{\bf I love Eric Clapton so much, words cannot describe it.} Eric Clapton - Siiiick BLUES SOLO!!! \#INSANE GUITARSOUND. [Video].Youtube. Retrieved July 7, 2021, from
\href{https://youtu.be/MLz2QmfUakY}{https://youtu.be/MLz2QmfUakY}.

\paragraph{S10 Video.}
\label{S10_Video}
{\bf DEF LEPPARD.} Phil Collen Plays ``White Lightning'' Solo for ``Adrenalize'' 20th. [Video].Youtube.
Retrieved July 7, 2021, from \href{https://youtu.be/FElcI66svKA}{https://youtu.be/FElcI66svKA}.

\paragraph{S11 Video.}
\label{S11_Video}
{\bf Henrik Danhage.} Evergrey Soaked Solo. [Video].Youtube.
Retrieved July 7, 2021, from
\href{https://youtu.be/hN__F0E-SFQ}{https://youtu.be/hN\_\_F0E-SFQ}.

\paragraph{S12 Video.}
\label{S12_Video}
{\bf Cornel Lazar.} U2's The Edge soundchecks his guitar rig (It Might Get Loud). [Video].Youtube.
Retrieved July 7, 2021, from
\href{https://youtu.be/xMw8NjCs_dghttps://youtu.be/xMw8NjCs_dg}{https://youtu.be/xMw8NjCs\_dg}.

\paragraph{S13 Video.}
\label{S13_Video}
{\bf Sweetwater.} Solos and Licks with Judas Priest's Richie Faulkner. [Video].Youtube.
Retrieved July 7, 2021, from
\href{https://youtu.be/p-UAqq8Bwbk}{https://youtu.be/p-UAqq8Bwbk}.

\paragraph{S14 Video.}
\label{S14_Video}
{\bf Orfeo Pomp.} John Frusciante Lesson with Under the Bridge. [Video].Youtube.
Retrieved July 7, 2021, from
\href{https://youtu.be/ILvNcif_f3E}{https://youtu.be/ILvNcif\_f3E}.

\paragraph{S15 Video.}
\label{S15_Video}
{\bf jlsrv.} David Gilmour - Comfortably Numb - AOL Music Sessions. [Video].Youtube.
Retrieved July 7, 2021, from
\href{https://youtu.be/Fw_9Xgv_oS0}{https://youtu.be/Fw\_9Xgv\_oS0}.

\paragraph{S16 Video.}
\label{S16_Video}
{\bf angelicproductions.} Billy Gibbons, Aleksanterinkatu, Helsinki, Finland. [Video].Youtube.
Retrieved August 22, 2021, from
\href{https://youtu.be/YHUQNxggT_k}{https://youtu.be/YHUQNxggT\_k}.

\paragraph{S17 Video.}
\label{S17_Video}
{\bf DLD2 Music!} Something (Isolated Guitar Solo). [Video].Youtube.
Retrieved August 22, 2021, from \\
\href{https://youtu.be/NZYItIFviM4}{https://youtu.be/NZYItIFviM4}.

\paragraph{S18 Video.}
\label{S18_Video}
{\bf Sd D.} Justin Hawkins Riding His White Tiger LIVE at Wembley Area. [Video].Youtube.
Retrieved July 7, 2021, from
\href{https://youtu.be/McUI5VwFbgY}{https://youtu.be/McUI5VwFbgY}.

\paragraph{S19 Video.}
\label{S19_Video}
{\bf Zulu Crunchbunch \& the Unstables.} Jimi Hendrix - Little Wing live in Frankfurt 1969. [Video].Youtube.
Retrieved July 7, 2021, from
\href{https://youtu.be/R2MI0tTRKfs}{https://youtu.be/R2MI0tTRKfs}.

\paragraph{S20 Video.}
\label{S20_Video}
{\bf Scotti Hill.} Scotti Hill I Remember You Solo Lesson. [Video].Youtube.
Retrieved July 7, 2021, from
\href{https://youtu.be/uM8JVFI3CTQ}{https://youtu.be/uM8JVFI3CTQ}.

\paragraph{S21 Video.}
\label{S21_Video}
{\bf CesarHuescaMusic.} Mi solo de guitarra para "RECUENTO FINAL", tema de Christianvib (PRONTO EN CALCANDO SOLOS). [Video].Youtube. Retrieved July 7, 2021, from
\href{https://youtu.be/d4Q7kj_4H4Y}{https://youtu.be/d4Q7kj\_4H4Y}.

\paragraph{S22 Video.}
\label{S22_Video}
{\bf Pavel Voronin.} Tony Iommi - The Riff Maker 1984 (eng). [Video].Youtube.
Retrieved July 7, 2021, from
\href{https://youtu.be/hcTE9wDTWZw}{https://youtu.be/hcTE9wDTWZw}.

\paragraph{S23 Video.}
\label{S23_Video}
{\bf John White.} Eric Johnson Clean Tone Signature. [Video].Youtube.
Retrieved July 7, 2021, from \\
\href{https://youtu.be/7u8jFYy7Rjs}{https://youtu.be/7u8jFYy7Rjs}.

\paragraph{S24 Video.}
\label{S24_Video}
{\bf lufunkClass.} KEZIAH JONES - LÂG S1000KJ SIGNATURE RUGGED 1 - DEMO PART 1. [Video].Youtube. Retrieved July 7, 2021, from
\href{https://youtu.be/TKPHnovuncY}{https://youtu.be/TKPHnovuncY}.

\paragraph{S25 Video.}
\label{S25_Video}
{\bf Arthur B.} Dire Straits - Sultans of Swing (Live, 1978). [Video].Youtube.
Retrieved July 7, 2021, from
\href{https://youtu.be/Z4Hzjca9rkk}{https://youtu.be/Z4Hzjca9rkk}.

\paragraph{S26 Video.}
\label{S26_Video}
{\bf .MARLBOROMAN55JP} Richie Kotzen YG Oct.2009 2/2. [Video].Youtube.
Retrieved July 7, 2021, from
\href{https://youtu.be/Ufk2jQUlSFc}{https://youtu.be/Ufk2jQUlSFc}.

\paragraph{S27 Video.}
\label{S27_Video}
{\bf davidwrightatloppers} Les Paul LIVE - and magnificent!. [Video].Youtube.
Retrieved August 22, 2021, from
\href{https://youtu.be/ZJaNMZmBR6E}{https://youtu.be/ZJaNMZmBR6E}

\paragraph{S28 Video.}
\label{S28_Video}
{\bf LarryDiMarzio.} DiMarzio Transition Guitar Pickups for Steve Lukather. [Video].Youtube.
Retrieved July 7, 2021, from
\href{https://youtu.be/9bmVt9k3zz4}{https://youtu.be/9bmVt9k3zz4}.

\paragraph{S29 Video.}
\label{S29_Video}
{\bf Official Yngwie Malmsteen.} Yngwie Malmsteen - Arpeggios from Hell. [Video].Youtube.
Retrieved July 7, 2021, from
\href{https://youtu.be/n6kPfe9gVJk}{https://youtu.be/n6kPfe9gVJk}.

\paragraph{S30 Video.}
\label{S30_Video}
{\bf smashbro.} Alejandro Marcovich tocando un solo extendida de "Afuera". [Video].Youtube.
Retrieved July 7, 2021, from
\href{https://youtu.be/9oh_wjaDp74}{https://youtu.be/9oh\_wjaDp74}.

\paragraph{S31 Video.}
\label{S31_Video}
{\bf Arte de guitarra.} QUEEN Bohemian Rhapsody (Únicamente guitarra/Guitar only) Brian May ORIGINAL RECORDING TRACK. [Video].Youtube. Retrieved July 7, 2021, from
\href{https://youtu.be/2jhxAbbnAQM}{https://youtu.be/2jhxAbbnAQM}.

\paragraph{S32 Video.}
\label{S32_Video}
{\bf Simon Nackborn.} John Mayer Improvising To Spotify Jam Tracks (IG Live, Sept. 25 2019). [Video].Youtube. Retrieved July 7, 2021, from \href{https://youtu.be/05X8X8RdQ2I}{https://youtu.be/05X8X8RdQ2I}.

\paragraph{S33 Video.}
\label{S33_Video}
{\bf Connecting to Cure Crohn's and Colitis.} Mike McCready from our 2020 Virtual Event. [Video].Youtube.
Retrieved July 7, 2021, from
\href{https://youtu.be/JrTUBirlYoQ}{https://youtu.be/JrTUBirlYoQ}.

\paragraph{S34 Video.}
\label{S34_Video}
{\bf LarryDiMarzio.} Gretchen Menn for DiMarzio PAF 59 Neck and Bridge pickups (4K). [Video].Youtube.
Retrieved July 7, 2021, from
\href{https://youtu.be/J7b9eRmopRg}{https://youtu.be/J7b9eRmopRg}.

\paragraph{S35 Video.}
\label{S35_Video}
{\bf Best Classic Bands.} Jimmy Page, Steve Miller Watch Don Felder Perform at Met Museum NYC 4/1/19. [Video].Youtube. Retrieved July 7, 2021, from
\href{https://youtu.be/14mgC1_Php8}{https://youtu.be/14mgC1\_Php8}.

\paragraph{S36 Video.}
\label{S36_Video}
{\bf John White.} Gary Moore clean jazz guitar solo. [Video].Youtube.
Retrieved July 7, 2021, from \\
\href{https://youtu.be/pC9FdvlWEkw}{https://youtu.be/pC9FdvlWEkw}.

\paragraph{S37 Video.}
\label{S37_Video}
{\bf Fender.} Tom Morello Fender Signature Sessions Fender. [Video].Youtube.
Retrieved July 7, 2021, from
\href{https://youtu.be/EZjtl8jGdac}{https://youtu.be/EZjtl8jGdac}.

\paragraph{S38 Video.}
\label{S38_Video}
{\bf Guitar Gear.} Galo Ochoa (Cuca/Nata) - ENGL Savage 120 Pt1. [Video].Youtube.
Retrieved July 7, 2021, from
\href{https://youtu.be/C9HQU61zb84}{https://youtu.be/C9HQU61zb84}.

\paragraph{S39 Video.}
\label{S39_Video}
{\bf jouvax.} orianthi solo. [Video].Youtube.
Retrieved July 7, 2021, from
\href{https://youtu.be/C3TCvnCDn1w}{https://youtu.be/C3TCvnCDn1w}.

\paragraph{S40 Video.}
\label{S40_Video}
{\bf emamele.} jimmy page's best solo. [Video].Youtube. Retrieved July 7, 2021, from
\href{https://youtu.be/PgA76eq2RTU}{https://youtu.be/PgA76eq2RTU}.

\paragraph{S41 Video.}
\label{S41_Video}
{\bf DT2BLZ.} 13 Minutes of Petrucci playing the guitar. [Video].Youtube.
Retrieved July 7, 2021, from \\
\href{https://youtu.be/iTOdwxdAkXY}{https://youtu.be/iTOdwxdAkXY}.

\paragraph{S42 Video.}
\label{S42_Video}
{\bf Elias Moisés.} Prince - Purple Rain (Guitar Solo in HD). [Video].Youtube.
Retrieved July 7, 2021, from
\href{https://youtu.be/JPGVD7Yy3Ys}{https://youtu.be/JPGVD7Yy3Ys}.

\paragraph{S43 Video.}
\label{S43_Video}
{\bf Guitarmaster2000.} Randy Rhoads - Guitar solo live with Quit Riot. [Video].Youtube.
Retrieved July 7, 2021, from
\href{https://youtu.be/HS8c0DWqIEI}{https://youtu.be/HS8c0DWqIEI}.

\paragraph{S44 Video.}
\label{S44_Video}
{\bf elnegrovilla.} The Rolling Stones - Sympathy For The Devil (Live at Tokyo Dome 1990). [Video].Youtube.
Retrieved August 22, 2021, from
\href{https://youtu.be/TAP1GHNuwQ8}{https://youtu.be/TAP1GHNuwQ8}.

\paragraph{S45 Video.}
\label{S45_Video}
{\bf Normans Rare Guitars.} Craig Ross from Lenny Kravitz's Band playing a 1964 Gibson ES-335TDC at Norman's Rare Guitars. [Video].Youtube. Retrieved July 7, 2021, from
\href{https://youtu.be/76g0VwCb1GM}{https://youtu.be/76g0VwCb1GM}.

\paragraph{S46 Video.}
\label{S46_Video}
{\bf TonioRuiz.} New Schecter TR-6 my signature model.. [Video].Youtube. Retrieved July 7, 2021, from
\href{https://youtu.be/l3JaGM-LEQM}{https://youtu.be/l3JaGM-LEQM}.

\paragraph{S47 Video.}
\label{S47_Video}
{\bf Bon Jovi Forever.} Richie Sambora Guitar Solo 1985. [Video].Youtube. Retrieved July 7, 2021, from
\href{https://youtu.be/7q4OuZoqRhA}{https://youtu.be/7q4OuZoqRhA}.

\paragraph{S48 Video.}
\label{S48_Video}
{\bf EpicConcerts.} Santana - Europa Live In London 1976. [Video].Youtube.
Retrieved July 7, 2021, from
\href{https://youtu.be/SVI7ZDDQXKA}{https://youtu.be/SVI7ZDDQXKA}.

\paragraph{S49 Video.}
\label{S49_Video}
{\bf Guitar Center.} Joe Satriani "Always With Me, Always With You" At: Guitar Center. [Video].Youtube.
Retrieved July 7, 2021, from
\href{https://youtu.be/czHYYVyBqCc}{https://youtu.be/czHYYVyBqCc}.

\paragraph{S50 Video.}
\label{S50_Video}
{\bf Moshcam.} Slash ft.Myles Kennedy \& The Conspirators - Blues Jam/Godfather Theme | Live in Sydney | Moshcam. [Video].Youtube. Retrieved July 7, 2021, from
\href{https://youtu.be/Fykcn5d7D6A}{https://youtu.be/Fykcn5d7D6A}.

\paragraph{S51 Video.}
\label{S51_Video}
{\bf torchytompsonisback.} Classic Albums The Number Of The Beast Extras (7).mp4. [Video].Youtube.
Retrieved July 7, 2021, from
\href{https://youtu.be/kxBjUZ70l_s}{https://youtu.be/kxBjUZ70l\_s}.

\paragraph{S52 Video.}
\label{S52_Video}
{\bf Steve Stine Guitar Lessons.} Steve Stine - Guitar Solo - Pink Floyd Inspired Jam. [Video].Youtube.
Retrieved July 7, 2021, from
\href{https://youtu.be/jI0tbO97bPc}{https://youtu.be/jI0tbO97bPc}.

\paragraph{S53 Video.}
\label{S53_Video}
{\bf kathy kielar.} Andy Summers 3-17. [Video].Youtube. Retrieved July 7, 2021, from
\href{https://youtu.be/RsN6CrUL4dQ}{https://youtu.be/RsN6CrUL4dQ}.

\paragraph{S54 Video.}
\label{S54_Video}
{\bf neveratime.} Pete Townshend - Magic Bus (Live) 1996. [Video].Youtube. Retrieved August 22, 2021, from
\href{https://youtu.be/pxZ77w9bnm8}{https://youtu.be/pxZ77w9bnm8}.

\paragraph{S55 Video.}
\label{S55_Video}
{\bf LarryDiMarzio.} Steve Vai for DiMarzio Dark Matter 2 Pickups. [Video].Youtube.
Retrieved July 7, 2021, from
\href{https://youtu.be/rLD00m9jRtQ}{https://youtu.be/rLD00m9jRtQ}.

\paragraph{S56 Video.}
\label{S56_Video}
{\bf Stevie Ray Vaughan.} Stevie Ray Vaughan \& Double Trouble - Pride And Joy (Live at Montreux 1982). [Video].Youtube. Retrieved July 7, 2021, from
\href{https://youtu.be/kfjXp4KTTY8}{https://youtu.be/kfjXp4KTTY8}.

\paragraph{S57 Video.}
\label{S57_Video}
{\bf Van Halen on MV.} Van Halen - Full Concert - 08/19/95 - Toronto (OFFICIAL). [Video].Youtube.
Retrieved July 7, 2021, from
\href{https://youtu.be/WFgkhz3-Dp0}{https://youtu.be/WFgkhz3-Dp0}.

\paragraph{S58 Video.}
\label{S58_Video}
{\bf LarryDiMarzio.} Angel Vivaldi demos DiMarzio Air Norton 7 and The Tone Zone 7 pickups. (4K). [Video].Youtube. Retrieved July 7, 2021, from
\href{https://youtu.be/hjVXsrv6P-0}{https://youtu.be/hjVXsrv6P-0}.

\paragraph{S59 Video.}
\label{S59_Video}
{\bf HiACDC.} Angus Young Guitar Solo. [Video].Youtube. Retrieved July 7, 2021, from
\href{https://youtu.be/en7EKL1pX5w}{https://youtu.be/en7EKL1pX5w}.

\paragraph{S60 Video.}
\label{S60_Video}
{\bf Dave David.} Frank Zappa - Black Napkins Oct.28, 1976. [Video].Youtube. Retrieved August 22, 2021, from
\href{https://youtu.be/_q0nImsfMvE}{https://youtu.be/\_q0nImsfMvE}.

\subsection*{Tables}

\begin{table}[ht]
\centering
\begin{tabular}{|l|l|l|l|}
\hline
\textbf{Name}	& \textbf{Time}	& \textbf{Time range} & \textbf{Reference}\\
\hline
$arpeggio$ & 0:06 & 0:00 - 0:06 & \nameref{S1_Audio} \\
\hline
$Gaussian\; noise$	& 0:06 & 0:00 - 0:06 & \nameref{S2_Audio} \\
\hline
\end{tabular}
\caption{\label{tableArpStat}{\bf Audio signals}. Type of signal, total time, and time interval.}
\end{table}

\begin{table}[ht]
\centering
\begin{tabular}{|l|l|l|l|l|}
\hline
\textbf{Name}	& \textbf{Time}	& \textbf{Time range} & \textbf{Reference}\\
\hline
$BBKing$*	& 1:44 & 1:45 - 3:28 & \nameref{S2_Video} \\
\hline
$Beck$	& 1:23 & 0:00 - 1:23 &  \nameref{S3_Video}\\
\hline
$Blackmore$	& 1:41 & 0:00 - 1:42 & \nameref{S5_Video} \\
\hline
$Bonamassa$	& 0:52 & 0:01 - 0:53 & \nameref{S6_Video} \\
\hline
$Clapton$	* & 1:16 & 0:00 - 1:16 & \nameref{S9_Video}  \\
\hline
$Edge$	& 1:07 & 0:00 - 1:09 & \nameref{S12_Video} \\
\hline
$Frusciante$	& 0:43 & 14:37 - 15:20 & \nameref{S14_Video} \\
\hline
$Gibbons$*	& 1:41 & 0:00 - 1:41 &  \nameref{S16_Video}\\
\hline
$Gilmour$	* & 2:31 & 4:36 - 7:05 & \nameref{S15_Video} \\
\hline
$Harrison$	& 0:34 & 0:00 - 0:34 &  \nameref{S17_Video}\\
\hline
$Hendrix$	& 1:49 & 2:02 - 3:52 & \nameref{S19_Video} \\
\hline
$Iommi$	& 1:12 & 22:15 - 23:27 & \nameref{S22_Video} \\
\hline
$Knopfler$*	& 0:54 & 4:08 - 5:04 & \nameref{S25_Video} \\
\hline
$Les Paul$	& 1:29 & 0:03 - 1:32 &  \nameref{S27_Video}\\
\hline
$Malmsteen$*	& 2:08 & 0:33 - 2:44 & \nameref{S29_Video} \\
\hline
$May$	& 0:42 & 0:44 - 1:26 & \nameref{S31_Video} \\
\hline
$Morello$	& 1:06 & 1:17 - 2:28 & \nameref{S37_Video} \\
\hline
$Page$	& 3:18 & 0:07 - 3:25 & \nameref{S40_Video} \\
\hline
$Prince$	& 1:54 & 0:04 - 2:08 & \nameref{S42_Video} \\
\hline
$Rhoads$	& 1:23 & 0:28 - 1:51 & \nameref{S43_Video} \\
\hline
$Richards$*	& 1:09 & 2:31 - 3:40 & \nameref{S44_Video} \\
\hline
$Santana$	& 1:09 & 0:05 - 1:20 & \nameref{S48_Video} \\
\hline
$Slash$	& 0:36 & 0:01 - 0:36 & \nameref{S50_Video} \\
\hline
$Summers$	& 2:28 & 0:06 - 2:39 & \nameref{S53_Video} \\
\hline
$Townshend$ & 0:28 & 1:10 - 1:40 &  \nameref{S53_Video} \\
\hline
$Vai$	& 0:57 & 0:00 - 0:57 & \nameref{S55_Video} \\
\hline
$Vaughan$*	& 1:07 & 1:36 - 2:47 & \nameref{S56_Video} \\
\hline
$VH$	& 2:26 & 3:15 - 5:41 & \nameref{S57_Video} \\
\hline
$Young$	& 0:55 & 0:00 - 0:56 & \nameref{S59_Video} \\
\hline
$Zappa$	& 1:21 & 0:45 - 2:01 & \nameref{S60_Video} \\
\hline
\end{tabular}
\caption{\label{tableTrial}{\bf Trial group of guitar players and their signal}. Name related to the the guitar player, total time, and time interval. Names with * indicates the application of a noise reduction in the file. There are 30 guitarist signals. See the list of videos in \href{https://youtube.com/playlist?list=PLesSAX_bg1DX4U1gPR_bKTbM3QMkMyQQ-}{Virtuosity (trial group)}.}
\end{table}

\begin{table}[ht]
\centering
\begin{tabular}{|l|l|l|l|l|}
\hline
\textbf{Name}	& \textbf{Time}	& \textbf{Time range} & \textbf{Reference}\\
\hline
$Batten$	&  0:45 & 0:01 - 0:46 & \nameref{S1_Video} \\
\hline
$Bettancourt$	& 2:44 & 7:39 -10:20 & \nameref{S4_Video} \\
\hline
$Borland$	& 1:33 & 3:44 - 5:16 & \nameref{S7_Video} \\
\hline
$Cerati$*	& 1:16 & 7:45 - 9:02 & \nameref{S8_Video} \\
\hline
$Collen$	& 1:07 & 0:26 - 1:33 & \nameref{S10_Video} \\
\hline
$Danhage$	& 0:35 & 0:00 - 0’:37 & \nameref{S11_Video} \\
\hline
$Faulkner$	& 0:43 & 0:02 - 0:43 & \nameref{S13_Video}  \\
\hline
$Hawkins$	& 0:56 & 0:25 - 1:27 & \nameref{S18_Video} \\
\hline
$Hill$	& 0:46 & 0:00 - 0:47 & \nameref{S20_Video} \\
\hline
$Huesca$*	& 0:50 & 0:05 - 0:58 & \nameref{S21_Video} \\
\hline
$Johnson$	& 1:00 & 0:00 - 0:58 & \nameref{S23_Video} \\
\hline
$Jones$	& 0:59 & 7:00 - 8:00 & \nameref{S24_Video} \\
\hline
$Kotzen$	& 0:38 & 0:05 - 0:43 & \nameref{S26_Video} \\
\hline
$Lukather$	& 1:01 & 0:00 - 1:01 & \nameref{S28_Video} \\
\hline
$Marcovich$*	& 3:03 & 0:01 - 3:00 & \nameref{S30_Video} \\
\hline
$Mayer$*	& 2:18 & 1:13 - 3:30 & \nameref{S32_Video} \\
\hline
$McCready$	& 1:05 & 2:53 - 4:00 & \nameref{S33_Video} \\
\hline
$Menn$	& 1:25 & 0:00 - 1:23 & \nameref{S34_Video} \\
\hline
$Miller$	& 2:10 & 0:45 - 2:57 & \nameref{S35_Video} \\
\hline
$Moore$	& 0:44 & 0:00 - 0:43 & \nameref{S36_Video} \\
\hline
$Ochoa$	& 1:07 & 3:58 - 5:05 & \nameref{S38_Video} \\
\hline
$Orianthi$	 &  0:57 & 0:00 - 0:57 & \nameref{S39_Video} \\
\hline
$Petrucci$	& 1:51 & 0:04 - 1:54 & \nameref{S41_Video} \\
\hline
$Ross$	& 2:20 & 0:02 - 2:22 & \nameref{S45_Video} \\
\hline
$Ruiz$	& 1:18 & 0:11 - 1:28 & \nameref{S46_Video} \\
\hline
$Sambora$	& 1:25 & 0:12 - 1:40 & \nameref{S47_Video} \\
\hline
$Satriani$	& 3:08 & 0:41 - 3:49 & \nameref{S49_Video} \\
\hline
$Smith$	& 1:27 & 0:00 - 1:28 & \nameref{S51_Video} \\
\hline
$Stine$*	& 1:42 & 0:06 - 1:48 & \nameref{S52_Video} \\
\hline
$Vivaldi$	& 1:26 & 0:00 - 1:26 & \nameref{S58_Video} \\
\hline
\end{tabular}
\caption{\label{tableControl}{\bf Control group of guitar players and their signal}. Name related to the the guitar player, total time, and time interval. Names with * indicates the application of a noise reduction in the file. There are 30 guitarist signals. See the list of videos in \href{https://youtube.com/playlist?list=PLesSAX_bg1DX4U1gPR_bKTbM3QMkMyQQ-}{Virtuosity (trial group)}.}
\end{table}

%%%%%%Audio
\begin{table}[ht]
\centering
\begin{tabular}{|l|l|l|l|l|}
\hline
{\bf Name} & {\bf Best and second best fit} & {\bf Parameters (a, b, loc, scale)} & {\bf KS test (d, p-value)}\\ 
\hline
$arpeggio$ & beta & (0.6343878714503692, & (0.0017292383591391292, \\
	&  & 18832.451641944805, & 0.8420322257740229) \\
	&  & 0.0024628816310980507, &  \\
	&  & 57117.37188806775) &  \\
	& exponential * & (0.002462881631098051, &(0.0018022684514384102, \\
	&  & 10.919175879654425) & 0.8038365894695755) \\
\hline

$Gaussian\; noise$ & Pareto * & (140.27760901257852, & (0.0013418394728997862, \\
	&  & -36152.58371514967, & 0.9710666454271314) \\
	&  & 36153.043244962406) &  \\
	& Weibull minimum & (0.17186145733895136, & (0.0015701121648797056, \\
	&  & 0.4954204131635887, & 0.9000840456757153) \\
	&  & 2.2637550911937083) &  \\
\hline
\end{tabular}
\caption{\label{tableBFArpStat}{\bf Audio files: Signals, best fit, and KS test.}. Name of the audio files, the best and second best fit statistical distribution, the estimated parameters of the KS goodness-of-fit test, and the KS test two-sided statistic. * Statistics of the selected best fit test.}
\end{table}

%%%Trial
\begin{table}[ht]
\centering
\begin{tabular}{|l|l|l|l|l|}
\hline
{\bf Name} & {\bf Best and second best fit} & {\bf Parameters (a, b, loc, scale)} & {\bf KS test (d, p-value)}\\ 
\hline
$BBKing$ & beta & (0.6941041242836588,  & (0.00034310184013075506,  \\
	&  & 1362.9285931299887, & 0.9509315437299324) \\
	&  &0.0005459481886498433 & \\
	&  & 57828.01992305121) & \\
	& exponential * & (0.0005459481886498435,  & (0.00037820815909361816, \\
	&  & 100.44777306037669) & 0.8996719660594641) \\
\hline

$Beck$ & normal & (85.94274007924642, & (0.0006134052855735206, \\
	&  & 269.0788909641859) &  0.4954541891665186)\\
	& lognormal * & (2.6275949223661406, & (0.0006134052855735206, \\
	&  & 0.0009408869160399091, &  0.4954541891665186)\\
	&  & 3.6292405056369703) &  \\
\hline

$Blackmore$ &  exponentiated Weibull *  & (1.3113847227920843, & (0.00040585860908165516, \\
	&  &  0.449350733642315, & 0.8558262673171991) \\
	&  & 0.010344087180073323, &  \\
	&  & 67.20125556818532) &  \\
	& Pareto & (0.3612635136755745, & (0.0004058586090817107, \\
	&  & -2.7408803175833745, & 0.8558262673170866) \\
	&  & 2.751224404548476) &  \\
\hline

$Bonamassa$ & exponential * & (4.958149709376698e-05, & (0.0007059166545807516,  \\
	&  & 77.70861221916532,  & 0.6145822128066771) \\
	& Pareto & (0.7539641353012134, & (0.0007059166545807516,  \\
	&  & -2.0706067611931402, & 0.6145822128066771) \\
	&  &  2.0706563426795297)& \\
\hline

$Clapton$ & normal & (57.98141907871804, & (0.000512581243781951, \\
	&  & 148.37166327619832) & 0.7714667514564951) \\
	& lognormal * & (2.3234130363246805, & (0.000512581243781951, \\
	&  & 0.0008843639527788647, & 0.7714667514564951) \\
	&  & 6.595573513561256) &  \\
\hline

$Edge$ & normal & (98.09924054320703, & (0.0005713248345645683, \\
	&  & 271.16806714843244) & 0.7184927293528525) \\
	& lognormal * & (2.647627306856628, & (0.0005713248345645683, \\
	&  & 0.00044334932372209134, & 0.7184927293528525) \\
	&  & 5.819964913867127) &  \\
\hline

$Frusciante$ & beta * & (0.3198717028450271, & (0.0006028934676896996, \\
	&  & 43.94048874698538, & 0.8819043726818238) \\
	&  & 0.006598469215286868, &  \\
	&  & 21860.18013914077) &  \\
	& normal & (154.74222325929398, & (0.0006339009658872241, \\
	&  & 441.75852587401573) & 0.8417962772923235) \\
\hline

$Gibbons$ & exponential * & (0.0021760047665896408, & (0.00040947589883566504, \\
	&  & 124.92284116236364) &  0.8503771584708064)\\
	& Weibull minimum & (0.5739328488430278, & (0.00040947589883566504, \\
	&  & 0.0021760047665896403, & 0.8503771584708064)\\
	&  & 70.58251317072336) &  \\
\hline
\end{tabular}
\caption{\label{tableBFTrial}{\bf Trial group: Signals, best fit, and KS test}. Name related to the the guitar player, the best and second best fit statistical distribution, the estimated parameters of the KS goodness-of-fit test, and the KS test two-sided statistic. * Statistics of the selected best fit test.}
\end{table}

\begin{table}[ht]
\centering
\begin{tabular}{|l|l|l|l|l|}
\hline
{\bf Name} & {\bf Best and second best fit} & {\bf Parameters (a, b, loc, scale)} & {\bf KS test (d, p-value)}\\ 
\hline
$Gilmour$ & exponential * & (0.0007033105706182077, & (0.0004199983160778631, \\
	&  & 68.70628982890044) & 0.5998709522672948) \\
	& Weibull minimum & (0.43628615325113795, & (0.0004199983160778631, \\
	&  & 0.0007033105706182076, & 0.5998709522672948) \\
	&  & 129.62371756588365) &  \\
\hline

$Harrison$ & exponential * & (0.00036839028878403755, & (0.000766449900662658, \\
	&  & 55.5665686536348) &  0.7746599552402547)\\
	& Pareto & (0.7819223662031394, & (0.000766449900662658, \\
	&  & -1.5197561242171502, & 0.7746599552402547) \\
	&  & 1.5201245144959183) &  \\
\hline

$Hendrix$ & beta * & (0.35780979655658146, & (0.0005088896830778289, \\
	&  & 475.42790384527456, & 0.5597948363204839) \\
	&  & 0.00024571482777053057, &  \\
	&  & 14002.1917153505) &  \\
	& normal & (58.0102181358127, & (0.0005255792997851949, \\
	&  & 220.90698062618284) & 0.5177859338678796) \\
\hline

$Iommi$ & gamma * & (0.21203091208445013, & (0.0005643725068888716, \\
	&  & 9.451147424339644e-05, & 0.6949366337090226) \\
	&  & 693.1832505445063) &  \\
	& normal & (76.56749271522027, & (0.0005810418712431931, \\
	&  & 172.79576599295072) & 0.6596707872556908) \\
\hline

$Knopfler$	& uniform & (0.00011896757803885322, & (0.000536076095170479, \\
	&  & 3097.5530946275876) & 0.8806328319148402) \\
	& exponential * & (0.00011896757803885322, & (0.0005360760951705901, \\
	&  & 63.74921374815442) & 0.8806328319146881) \\
\hline

$Les Paul$	& exponentiated Weibull & (1.5582795837638401, & (0.00047100211573875006, \\
	&  & 0.33248826227216666, &  0.7769832611500577)\\
	&  & 0.0006962811101605195, &  \\
	&  & 5.76650181275355) &  \\
	& exponential * & (0.0006962811101605196, & (0.00047100211573880557, \\
	&  & 74.28158935398821) &  0.7769832611499349)\\
\hline

$Malmsteen$ & exponentiated Weibull & (1.1064981865786412, & (0.0003594215541037893, \\
	&  & 0.4310583919668194, & 0.8600428887373286) \\
	&  & 0.0013033554687252368, & \\
	&  & 48.217843804727906) &  \\
	& exponential * & (0.001303355468725237, & (0.00035942155410381704, \\
	&  & 146.58197072646385) & 0.8600428887372661) \\
\hline

$May$ 	& normal & (97.05126488867273, & (0.0006276133928297956, \\
	&  & 270.7051767117866) & 0.860382886438023) \\
	& lognormal * & (2.3306573757062514, & (0.0006276133928297956, \\
	&  & 0.0014685591246090808, & 0.860382886438023) \\
	&  & 7.307988525151062) &  \\
\hline
\end{tabular}
\caption*{\label{tableBFTrial0}{\bf Table 5. Trial group: Signals, best fit, and KS test (Continued)}. Name of the audio files, the best and second best fit statistical distribution, the estimated parameters of the KS goodness-of-fit test, and the KS test two-sided statistic. * Statistics of the selected best fit test.}
\end{table}

\begin{table}[ht]
\centering
\begin{tabular}{|l|l|l|l|l|}
\hline
{\bf Name} & {\bf Best and second best fit} & {\bf Parameters (a, b, loc, scale)} & {\bf KS test (d, p-value)}\\ 
\hline
$Morello$	& normal & (63.89957811393796, & (0.0005403450389701581, \\
	&  & 204.30853064964717) & 0.7904583507584694) \\
	& lognormal * & (2.979053342417809, & (0.0005403450389701581, \\
	&  & 0.0002348887986833976, & 0.7904583507584694) \\
	&  & 1.6268350034682917) &  \\

\hline

$Page$ & exponential * & (0.00011072561556800662, & (0.00036887474613900295, \\
	&  & 87.4120088858035) & 0.5931995888548788) \\
	& Pareto & (0.6807583019373199, & (0.00036887474613900295, \\
	&  & -2.402265574806937, & 0.5931995888548788) \\
	&  & 2.4023763004116137) &  \\
\hline

$Prince$ & exponential * & (0.0030883343858422656, & (0.000468003298502051, \\
	&  & 201.21468773856972) & 0.6381685675956954) \\
	& Pareto & (0.38794601255390004, & (0.000468003298502051, \\
	&  & -2.8000409689645034, & 0.6381685675956954) \\
	&  & 2.803129303237643) &  \\
\hline

$Rhoads$ & exponential * & (0.00043419218584243234, & (0.0006023639773738987, \\
	&  & 89.07852464659618) & 0.5164735691678086) \\
	& Pareto & (0.4214803841754926, & (0.0006023639773738987, \\
	&  & -0.7609832083259636, & 0.5164735691678086) \\
	&  & 0.7614174004872494) &  \\
\hline

$Richards$ & normal & (55.083231968337586, & (0.0005706843000541362, \\
	&  & 144.15196711142943) &  0.7125354610054572)\\
	& lognormal * & (2.509074988126237, & (0.0005706843000541362, \\
	&  & 0.0006749125051491495, & 0.7125354610054572) \\
	&  & 4.393323977381983) &  \\
\hline

$Santana$ & normal & (105.70092393097259, & (0.0006387205304270172, \\
	&  & 287.49059634356485) & 0.5656988523847863) \\
	& lognormal * & (2.070800110274306, & (0.0006387205304270172, \\
	&  & 0.0036925929638214703, & 0.5656988523847863) \\
	&  & 17.096193864510216) &  \\
\hline

$Slash$ & beta & (0.4491613291930282, & (0.0008140547108498875, \\
	&  & 50.26210915747921, & 0.6618601885010457) \\
	&  & 0.003993473933777688, &  \\
	&  & 3851.234642681255) &  \\
	& exponential * & (0.003993473933777717, & (0.0008855410222965432, \\
	&  & 69.16340873890937) & 0.5547912992547333) \\
\hline

$Summers$ & exponential * & (0.0006029648444517348, & (0.00037507592991159466, \\
	&  & 69.81824966238491) & 0.7490254603051878) \\
	& Weibull minimum & (0.4368509724647492, & (0.00037507592991159466, \\
	&  & 0.0006029648444517347, & 0.7490254603051878) \\
	&  & 15.723651816109506) &  \\
\hline

$Townshend$ & Gilbrat * & (-4.919100006015126, & (0.0008070338097219132, \\
	&  & 17.652471033098365) & 0.8201892516455872) \\
	& normal & (64.92777796605147, & (0.0008070338097220242, \\
	&  & 156.20342120158088) & 0.820189251645459) \\
\hline
\end{tabular}
\caption*{\label{tableBFTrial1}{\bf Table 5. Trial group: Signals, best fit, and KS test (Continued)}. Name related to the the guitar player, the best and second best fit statistical distribution, the estimated parameters of the KS goodness-of-fit test, and the KS test two-sided statistic. * Statistics of the selected best fit test.}
\end{table}

\begin{table}[ht]
\centering
\begin{tabular}{|l|l|l|l|l|}
\hline
{\bf Name} & {\bf Best and second best fit} & {\bf Parameters (a, b, loc, scale)} & {\bf KS test (d, p-value)}\\ 
\hline
$Vai$ & exponential *& (0.0006541804315897288, & (0.0005583312083495828, \\
	&  & 78.16285298630608, & 0.8277751756052413) \\
	& Pareto & (0.5526769614962832, & (0.0005583312083495828, \\
	&  & -1.6551712990641976, & 0.8277751756052413) \\
	&  & 1.6558254793970968) &  \\
\hline

$Vaughan$ & normal & (57.32195952794705, & (0.0005600170780110103, \\
	&  & 158.79560977147352) & 0.740693632464756) \\
	& lognormal * & (2.5393671568102225, & (0.0005600170780110103, \\
	&  & 0.0006956940669767824, & 0.740693632464756) \\
	&  & 4.760497389639097) &  \\
\hline

$VH$ & exponential * & (9.508796892115721e-05, & (0.00038937588165977033, \\
	&  & 44.3546208291969) & 0.7150967557576102) \\
	& exponentiated Weibull & (1.941514141613077, & (0.00038937588165977033, \\
	&  & 0.271927153972658, & 0.7150967557576102) \\
	&  & 9.50879689211572e-05, &  \\
	&  & 1.5603486025977524) &  \\
\hline

$Young$ & exponential *& (0.00045301322208451825, & (0.0005178227307015559, \\
	&  & 72.18367300089396, & 0.9009045727799112) \\
	& Pareto & (0.6161202344613161, & (0.0005178227307015559, \\
	&  & -1.525571883596792, & 0.9009045727799112) \\
	&  & 1.526024896796291) &  \\
\hline

$Zappa$ & Gilbrat * & (-5.428894664051688, & (0.0005892311156233676, \\
	&  & 23.031501752240608) & 0.5684587669641091) \\
	& normal & (86.30010382647654, & (0.0005892311156234786, \\
	&  & 242.7768683467798) & 0.5684587669638648) \\
\hline
\end{tabular}
\caption*{\label{tableBFTrial2}{\bf Table 5. Trial group: Signals, best fit, and KS test (Continued)}. Name of the audio files, the best and second best fit statistical distribution, the estimated parameters of the KS goodness-of-fit test, and the KS test two-sided statistic. * Statistics of the selected best fit test.}
\end{table}

\begin{table}[ht]
\centering
\begin{tabular}{|l|l|l|l|l|}
\hline
{\bf Name} & {\bf Best and second best fit} & {\bf Parameters (a, b, loc, scale)} & {\bf KS test (d, p-value)}\\ 
\hline
$Batten$ & beta ** & (0.3031944350259428,  & (0.0005341260060560682,  \\
	&  & 1644.4333575198198, & 0.9392030290092083) \\
	&  & 0.0005003350246539589, &  \\
	&  & 117213.98187040549) &  \\
	& normal & (57.88724739045192,  & (0.0006644684656098576,  \\
	&  & 144.52605882262068) & 0.7722860082941468) \\	
\hline

$Bettancourt$ & beta & (0.7636373319994032, & (0.0003588577572226359, \\
	&  & 71240578.43719171, & 0.7406612194051129) \\
	&  & 0.0021161259626267265, &  \\
	&  & 7014559275.663902) &  \\
	& expon * & (0.002117490831770411, & (0.00047014165782738315, \\
	&  & 206.46189237404644) & 0.4014508093885853) \\
\hline

$Borland$ & exponential * & (0.000242411272236367, & (0.00042544870109417765, \\
	&  & 92.05895433255284)  & 0.850256807809482) \\
	& Pareto & (0.6590829226672976,  & (0.00042544870109417765, \\
	&  & -4.518052618127804, & 0.850256807809482) \\
	&  & 4.518295029270276) &  \\
\hline

$Cerati$	& normal & (56.034458187682674, & (0.0005138575585692529, \\
	&  & 151.926737422471) & 0.7678011319060685) \\
	& lognormal * & (2.403459199542931, & (0.0005138575585692529, \\
	&  & 0.0005974832804988078, & 0.7678011319060685) \\
	&  & 5.720081701271518) &  \\
\hline

$Collen$ & normal & (165.5215125728059, & (0.0006073944350787341, \\
	&  & 406.22802238614685) & 0.6463943836469558) \\
	& lognormal * & (2.02915794123515, & (0.0006073944350787341, \\
	&  &0.0026922346568573566, & 0.6463943836469558) \\
	&  & 25.482516854669612) &  \\
\hline

$Danhage$ & beta & (0.4985000353539132, & (0.0006699871095987575, \\
	&  & 589.5898290632397, & 0.8757844762092282) \\
	&  & 0.0007539715836758074, &  \\
	&  & 16915.90300812343) &  \\
	& exponential * & (0.0007539715836758075, & (0.000797102703009811, \\
	&  & 61.18681348119881)  & 0.7057290422572016) \\
\hline

$Faulkner$ & normal & (114.73929911571895, & (0.0006315354725340283, \\
	&  & 265.45427385880134) & 0.8407710335082792) \\
	& lognormal * & (2.2623321061099837, & (0.0006315354725340283, \\
	&  & 0.0015405011042798516, & 0.8407710335082792) \\
	&  & 11.778672459454608) &  \\
\hline

$Hawkins$ & exponential & (0.0009079795915529996, & (0.0005024343704149725, \\
	&  & 79.01732608931998) & 0.9121022576882232) \\
	& exponentiated Weibull * & (1.6027517889529332, & (0.0005024343704149725, \\
	&  & 0.33482259203373943, & 0.9121022576882232) \\
	&  & 0.0009079795915529995, &  \\
	&  & 7.6564767296831135) &  \\
\hline
\end{tabular}
\caption{\label{tableBFControl}{\bf Control group: Signals, best fit, and KS test}. Name related to the the guitar player, the best and second best fit statistical distribution, the estimated parameters of the KS goodness-of-fit test, and the KS test two-sided statistic. * Statistics of the selected best fit test.}
\end{table}

\begin{table}[ht]
\centering
\begin{tabular}{|l|l|l|l|l|}
\hline
{\bf Name} & {\bf Best and second best fit} & {\bf Parameters (a, b, loc, scale)} & {\bf KS test (d, p-value)}\\ 
\hline
$Hill$	& beta ** & (0.40775450207110064, & (0.00048596794648758657, \\
	&  & 533.7271583805846, & 0.9712420088721049) \\
	&  & 0.0005742051023517377, &  \\
	&  & 42133.19793573992) &  \\
	& normal & (90.39553021986232, & (0.0007314844733120607, \\
	&  & 271.60615827722233) & 0.6538502326137732) \\
\hline

$Huesca$ & exponential * & (0.0005032928141684691, & (0.0006575362809206897, \\
	&  & 49.522201386626385) & 0.7269496513937523) \\
	& Pareto & (0.6189892508219239, & (0.0006575362809206897, \\
	&  & -2.2801291392389293, & 0.7269496513937523) \\
	&  & 2.2806324319177538) &  \\
\hline

$Johnson$ & normal & (67.95630261509292, & (0.000637554773848803, \\
	&  & 197.80627834491904) & 0.6567984373960525) \\
	& lognormal * & (2.118464345716879, & (0.000637554773848803, \\
	&  & 0.0004564221050809147, & 0.6567984373960525) \\
	&  & 7.512944220414095) &  \\
\hline

$Jones$ & beta **& (0.2763656493390366, & (0.0006255428765344906, \\
	&  & 307533.4685505172, & 0.6893656749028716) \\
	&  & 0.0006163018122106163, &  \\
	&  & 54770660.62121732) &  \\
	& normal & (53.816252889330684, & (0.0006326963572547695, \\
	&  & 133.43943207042653) & 0.6756577438659741) \\
\hline

$Kotzen$ & gamma * & (0.25419186644680763, & (0.0007118465326173551, \\
	&  & 0.002222846841806395, & 0.786903055212888) \\
	&  & 772.8009501948068) &  \\
	& beta & (0.6015122226548679, & (0.0008103031669770555, \\
	&  & 264.38047615125896, & 0.6378732020526754) \\
	&  & 0.002222846841806395, &  \\
	& & 11240.373491925788) &  \\
\hline

$Lukather$	& normal & (63.325810818903435, & (0.0006521678034260292, \\
	&  & 180.3459112993755) & 0.6193888951436818) \\
	& lognormal * & (2.69018630504822, & (0.0006521678034260292, \\
	&  & 0.0001820606279622696, & 0.6193888951436818) \\
	&  & 2.690435854698224) &  \\
\hline

$Marcovich$	& beta & (0.35583253664233916, & (0.00031575645329862123, \\
	&  & 97281821.95632735, & 0.8160965001085185) \\
	&  & 0.0002557579750628631, &  \\
	&  & 8730263589.309713) &  \\
	& exponential * & (0.00025575802028615614, & (0.0002704816963865764, \\
	&  & 75.57243793960556) & 0.9294562573705455) \\
\hline

$Mayer$ & beta & (0.2876421462441351, & (0.0003572764579702348, \\
	&  & 322.91779510260005) &0.8311793025780103) \\
	&  & 0.0014134299134735328, &  \\
	&  & 162306.10152241576) &  \\
	& exponential * & (0.001413429913473533, & (0.0003715689565381153, \\
	&  & 143.9129082358288) & 0.7938165218527682) \\
\hline

$McCready$ & normal & (89.23642366249419, & (0.0005673493865773205, \\
	&  & 247.34925967273867) & 0.7482195455738625) \\
	& lognormal * & (2.6233515366472018, & (0.0005673493865773205, \\
	&  & 0.0007910186709333624, & 0.7482195455738625) \\
	&  & 5.079173146693279) &  \\
\hline
\end{tabular}
\caption*{\label{tableBFControl0}{\bf Table 6. Control group: Signals, best fit, and KS test (Continued)}. Name of the audio files, the best and second best fit statistical distribution, the estimated parameters of the KS goodness-of-fit test, and the KS test two-sided statistic. * Statistics of the selected best fit test.}
\end{table}

\begin{table}[ht]
\centering
\begin{tabular}{|l|l|l|l|l|}
\hline
{\bf Name} & {\bf Best and second best fit} & {\bf Parameters (a, b, loc, scale)} & {\bf KS test (d, p-value)}\\ 
\hline
$Menn$	& exponential * & (0.0009313054847240391, & (0.0005530674632635568, \\
	&  & 54.88098484476828) & 0.6167909105778449) \\
	& Pareto & (0.5035028222132063, & (0.0005530674632635568, \\
	&  & -1.3704038532237983,  & 0.6167909105778449) \\
	&  & 1.3713351586781375) &  \\
\hline

$Miller$ & exponential * & (0.00031469542793777173, & (0.00034058835495254236, \\
	&  & 169.45890834066097) & 0.8934751009352255) \\
	& Pareto &(0.47773816905914457,  & (0.00034058835495254236, \\
	&  & -3.012893424239853, & 0.8934751009352255) \\
	&  & 3.013208119606605) &  \\
\hline

$Moore$ & norm & (70.19754045547499, & (0.00059168319241542649, \\
	&  & 231.76859309627352) & 0.886659506944159) \\
	& lognormal * & (2.134874037845794, & (0.0005916831924154264, \\
	&  & 0.0009964418478056376, & 0.886659506944159) \\
	&  & 6.224276227076958) &  \\
\hline

$Ochoa$	& normal & (88.6363474204818, & (0.0005859433748728327, \\
	&  & 244.86913753221705) & 0.6901189766470227) \\
	& lognormal * & (2.5667317094114237, & (0.0005859433748728327, \\
	&  & 0.0004208527218159566,  & 0.6901189766470227) \\
	&  & 4.691681043554068) &  \\
\hline

$Orianthi$ & exponential * & (0.0009149916009687099, & (0.0004977903184411048, \\
	&  & 84.57224676784833, & 0.9153603273147564) \\
	& Pareto & (0.5799983677418395, & (0.0004977903184411048, \\
	&  & -2.3104578907318816, & 0.9153603273147564) \\
	&  & 2.3113728822880297) &  \\
\hline

$Petrucci$ & normal & (129.9560072125629, & (0.0005420871231213709, \\
	&  & 316.540972918776, & 0.47087132345044264) \\
	& lognormal * & (3.0209441322750035, & (0.0005420871231213709, \\
	&  & 0.0003528549997325607, & 0.47087132345044264) \\
	&  & 5.47696892190317) &  \\
\hline

$Ross$	& beta & (0.7638041599373884, & (0.0003409735426713745, \\
	&  & 632.7466387963244, & 0.8751355872010046) \\
	&  & 0.007457677400764833, &  \\
	&  & 75674.67275919711) &  \\
	& exponential * & (0.007457677400764834, & (0.0003868043289453893, \\
	&  & 236.69299396660512) & 0.7453424486392385) \\
\hline

$Ruiz$ & normal & (67.90554918829253, & (0.0005589499697257194, \\
	&  & 214.16921868361408) & 0.659302481512031) \\
	& lognormal * & (2.596020512375068, & (0.0005589499697257194, \\
	&  & 0.00045958594744876266, & 0.659302481512031) \\
	&  & 4.390755231600053) &  \\
\hline
\end{tabular}
\caption*{\label{tableBFControl1}{\bf Table 6. Control group: Signals, best fit, and KS test (Continued)}. Name of the audio files, the best and second best fit statistical distribution, the estimated parameters of the KS goodness-of-fit test, and the KS test two-sided statistic. * Statistics of the selected best fit test.}
\end{table}

\begin{table}[ht]
\centering
\begin{tabular}{|l|l|l|l|l|}
\hline
{\bf Name} & {\bf Best and second best fit} & {\bf Parameters (a, b, loc, scale)} & {\bf KS test (d, p-value)}\\ 
\hline
$Sambora$ & beta & (0.2339690261901173, & (0.0004718113226664933, \\
	&  & 1642.3329719669973, & 0.7972989139061427) \\
	&  & 0.00033374296067277044, &  \\
	&  & 542344.003835276) &  \\
	& exponential * & (0.00033707658025589163, & (0.00054634961095823, \\
	&  & 76.7696398426536) & 0.6293194392181809) \\
\hline

$Satriani$ & beta & (0.32870110349493586, & (0.00031697513189044013, \\
	&  & 24.274370996438975, & 0.79889259739541) \\
	&  & 0.00024291369117680452, &  \\
	&  & 8606.60495410007) &  \\
	& exponential * & (0.000242913691176808, & (0.00040005221895345056, \\
	&  & 112.21804781121854) & 0.5203794697714443) \\
\hline

$Smith$ & beta & (0.4184269390567937, & (0.0005288759876340698, \\
	&  & 441.19139554940904, & 0.6532287392713398) \\
	&  & 0.006732342109275801, &  \\
	&  & 268926.6859855589) &  \\
	& exponential * & (0.006732342109275802, & (0.0005492466908365023, \\
	&  & 254.1847653986564) & 0.6056121680890219) \\
\hline

$Stine$	& exponential * & (7.437337124719251e-05, & (0.00043006392599992393, \\
	&  & 64.90345522604197) & 0.8015918345962205) \\
 	& Pareto & (0.7895769635444678, & (0.00043006392599992393, \\
	&  & -2.099949904457059, & 0.8015918345962205) \\
	&  & 2.100024277782845) &  \\
\hline

$Vivaldi$ & beta * & (0.2223861759027858, & (0.00047989357465000326, \\
	&  & 23.053368990357534, & 0.7764711002760281) \\
	&  & 0.00016257316368007362, &  \\
	&  & 7890.949530888383) &  \\
	& exponentiated Weibull  & (1.0698916435488428, & (0.0005445851878134178, \\
	&  & 0.327372429780207, & 0.6290303244896941) \\
	&  & 0.0001625731639892197, &  \\
	&  & 11.741661549302385) &  \\
\hline
\end{tabular}
\caption*{\label{tableBFControl2}{\bf Table 6. Control group: Signals, best fit, and KS test (Continued)}. Name of the audio files, the best and second best fit statistical distribution, the estimated parameters of the KS goodness-of-fit test, and the KS test two-sided statistic. * Statistics of the selected best fit test.}
\end{table}

\begin{table}[ht]
\centering
\begin{tabular}{|l|l|l|l|l|l|}
\hline
{\bf Group} & {\bf Best and} & {\bf Parameters (a, b, loc, scale)} & {\bf KS test (d, p-value)} &{\bf First moments} \\ 
 & {\bf second best fit} &  &  & \\ 
\hline
$Trial$ & beta & (0.9254129485162816, & (0.09180096752880187, & (0.544709475,\\
	&  & 0.8242795490069899, & 0.9621164168947198) & 0.53867906,\\
	&  & 0.2918501278078749, &  & 0.01973545,\\
	&  & 0.4666832656381216) &  & -0.10241556,\\
	&  &  &  & -1.25081877) \\
	&  &  &  & \\
	& gamma * & (122.44451502094967, & (0.12914869508242166, & (\textbf{0.5314125754},\\
	&  & -0.7485414426808097, & 0.6770027753472341) & 0.53490483,\\
	&  & 0.010481860092354429) &  & 0.0134529,\\
	&  &  &  & 0.18074252, \\
	&  &  &  &  0.04900179)\\
\hline
$Control$ & gamma * & (900.1898213776649, & (0.12909839636580123, & (\textbf{0.5129645989},\\
	&  & -2.5676631398085332, & 0.6775534048290307) & 0.51410568,\\
	&  & 0.0034234655242896435) &  & 0.01055033,\\
	&  &  &  & 0.06665964, \\
	&  &  &  & 0.00666526) \\
	&  &  &  & \\
	& lognormal & (0.0064590686756918845, & (0.15469363388940927, & (0.51408072,\\
	&  & -15.261998240919922, & 0.4303635601912797) & 0.51440981,\\
	&  & 15.77607896436885) &  & 0.01038401,\\
	&  &  &  & 0.01937768, \\
	&  &  &  & 0.00066755) \\
\hline
\end{tabular}
\caption{\label{bestFitMWUtFull}{\bf Diversity, best fit, KS test, and first moments}. Group is related to trial and control, the best and second best fit statistical distribution, the estimated parameters of the KS goodness-of-fit test, and the KS test two-sided statistic. * Statistics of the selected best fit test. First moments are related to the following statistics: (median, mean, variance, skewness, kurtosis). Bold values in the First moments column indicates the statistic of the median.}
\end{table}

\begin{table}[ht]
\centering
\begin{tabular}{|c|c|}
\hline
{\bf Name} & {\bf PDF}  \\
\hline
exponential	&  $f(x) = exp(-x)$, for $x >= 0$\\
gamma		& $f(x, \alpha) = \frac{x^{\alpha -1} e^{-x}}{\Gamma (\alpha)}$, for $x >=0$, $\alpha > 0$ \\
Gilbrat		&  $f(x)=\frac{1}{x \sqrt{2\pi}} exp(-\frac{1}{2}(log(x)^2))$\\
beta			&  $f(x,a,b)=\frac{\gamma(a+b)x^{\alpha -1}(1-x)^{b-1}}{\gamma (a) \gamma ()}$\\
exponentiated Weibull	& $f(x, \alpha, c) = \alpha c [1-exp(-x^{c})]^{\alpha -1} exp(-x^{c})x^{c-1}$, for $x > 0$, $\alpha > 0$, $c >0$ \\
lognormal		&  $f(x, s) = \frac{1}{sx \sqrt[]{2\pi}}exp(- \frac{log^{2}(x)}{2s^{2}})$, for $x > 0$, $s > 0$\\
\hline
\end{tabular}
\caption{\label{pdfs}{\bf Statistical distributions and their probability density functions (PDF).} }
\end{table}

\end{document}